\documentclass[11pt]{article}
\usepackage{amssymb}

\usepackage{graphicx}
\usepackage{amsmath}
\usepackage{makeidx}
\usepackage{indentfirst}

\newcounter{resultnum}[section]

\newcounter{conclusionnum}[section]

\newcounter{conditionnum}[section]

\newcounter{conjecturenum}[section]

\newcounter{examplenum}[section]

\newcounter{exercisenum}[section]

\newcounter{lemmanum}[section]

\newcounter{notationnum}[section]

\newcounter{theoremnum}[section]

\newcounter{definitionnum}[section]

\newcounter{corollarynum}[section]

\newcounter{remarknum}[section]

\newcounter{propositionnum}[section]

\newcounter{acknowledgementnum}[section]

\newcounter{algorithmnum}[section]

\newcounter{axiomnum}[section]

\newcounter{casenum}[section]

\newcounter{claimnum}[section]

\newcounter{summarynum}[section]

\newcounter{problemnum}[section]

\begin{document}

\title{New Classes of Off--Diagonal Cosmological Solutions in Einstein Gravity}
\date{June 16, 2010}
\author{Sergiu I. Vacaru\thanks{
sergiu.vacaru@uaic.ro, Sergiu.Vacaru@gmail.com;\newline
http://www.scribd.com/people/view/1455460-sergiu } \\
%EndAName
{\quad} \\
{\small {\textsl{\ Science Department, University "Al. I. Cuza" Ia\c si},} }%
\\
{\small {\textsl{\ 54 Lascar Catargi street, 700107, Ia\c si, Romania}} }}
\maketitle

\begin{abstract}
In this work, we apply the anholonomic deformation method for constructing new classes of anisotropic cosmological solutions in Einstein gravity and/or generalizations with nonholonomic variables. The\-re are analyzed four types of, in general, inhomogeneous metrics, defined with respect to anholonomic
frames and their main geometric properties. Such spacetimes contain as particular cases certain conformal and/or frame transforms of the well known Friedman--Robertson--Walker, Bianchi, Kasner and G\"{o}del universes and define a great variety of cosmological models with generic off--diagonal metrics, local anisotropy and inhomogeneity. It is shown that certain nonholonomic gravitational configurations may mimic de Sitter like inflation scenarios and different anisotropic modifications without satisfying any classical false--vacuum equation of state. Finally, we speculate on perspectives when such off--diagonal solutions can be related to dark energy
and dark matter problems in modern cosmology.

\vskip5pt


\vskip3pt


\end{abstract}

\tableofcontents

\bigskip

\newpage

\section{Introduction}

Modifications of general relativity (GR) theory and new classes of
cosmological solutions have received much attention as attempts to account
for dark energy and recent observations from the Wilkinson Microwave
Anisotropic Probe (WMAP), see \cite{wmap,lambda,planck,odintsov,sotr,capozz}
for reviews. There is certain evidence of relatively small anisotropic
departures from the standard Friedmann--Robertson--Walker (FRW) model.
However, we argue that it may be possible to involve more general classes of
anisotropic and/or inhomogeneous cosmological solutions, described by
generic off--diagonal metrics\footnote{%
such metrics can not be diagonalized by any coordinate transforms} in GR, in
order to explain in the bulk the existing experimental data and examine
various types of cosmological scenaria.

A series of Bianchi models with anisotropies has been analyzed. There is a
classification of Bianchi metrics arranging all possible spatially
homogeneous models depending on the symmetry properties of their spatial
hypersurfaces, see recent developments and references in \cite%
{calog,sungcoles1}. A few of Bianchi universes contain the FRW model as a
limiting case.\footnote{%
The paradigm of modern cosmology is based on the
Friedmann--Robertson--Walker (FRW) metric which is derived as a spherical
symmetric solution of the Einstein equations assuming homogeneity and
isotropy on large scales.} But even the best such models (the so--called
Bianchi $VII_{h}$ class) seem to be inconsistent \cite{pontzen} with WMAP
data.

Another classes of anisotropic and/or inhomogeneous solutions are used in
Kasner and G\"{o}del cosmological models \cite%
{frolov,roberts,steinh,godel,obukh1,obukhov,barrow}. Nevertheless, the
problem to construct the cosmological model and related solutions of
gravitational and matter field equations which would describe most
realistically the existing experimental data is still unsolved in modern
gravity, cosmology and astrophysics.

Recently, we developed the so--called anholonomic deformation method of
constructing exact solutions in gravity \cite{vgsolhd} (see examples and
reviews in \cite{vncg,vncbh,ijgmmp,vsgg,vrflg}). Perhaps, this is the unique
existing at present geometric method providing a formalism for generating
very general classes of solutions of gravitational field equations in GR and
various high dimension/ metric--affine, Lagrange--Finsler, noncommutative or
other extensions. The method is based on nonlinear connection geometry
originally elaborated for Finsler spaces, see details in \cite{ma} and, in
relation to standard theories of physics, in \cite{vrflg,ijgmmp}. In this
work, we shall not concern any questions related to details for scenaria of
"Finsler cosmology" but consider some geometric techniques for generating
cosmological solutions parametrized by generic off--diagonal metrics. The
goal of this paper is to construct and analyze new classes of cosmological
metrics for four dimensional, 4--d, pseudo--Riemannian spacetimes.

The off--diagonal anisotropic cosmological solutions to be considered in
this work are for GR. We shall use an auxiliary linear connection (the
so--called canonical distinguished connection, d--connection), for which the
Einstein equations can be solved in general form with respect to some
adapted nonholonomic frames of reference.\footnote{%
In modern literature on mathematics and physics, there are considered three
equivalent terms: nonholonomic, anholonomic and/or non--integrable (for
convenience, we shall use all such terms). Here, we also note that our
approach should not be confused with the so--called Cartan's moving frame
method when some geometric/ physical objects are re--defined with respect to
some more convenient frames of references/local basis. We consider
nonholonomic deformations of geometric/ physical objects (for instance,
deformations of the Levi--Civita connection) by imposing non--integrable
constraints on the dynamics of gravitational fields and anholonomic frames
with associated nonlinear connections structure (the last one being defined
as a conventional horizontal (h) and vertical (v) spacetime splitting).}
Such a connection is also metric compatible and completely defined by the
metric structure but contains nontrivial torsion terms induced
nonholonomically by some off--diagonal coefficients of metric. Having
constructed generalized exact (cosmological) solutions for an auxiliary
connection, we have to impose certain additional constraints on coefficients
of metrics in order to "extract" exact solutions for the Levi--Civita
connection.

We emphasize that the anholonomic deformation method allows us to generate
"almost all" classes of solutions when the time like coordinate is contained
as a "nonholonomic" one. Such generic off--diagonal solutions possess at
least a Killing vector symmetry and, in general, depend additionally on two
space coordinates. Various models of nonhomogeneous and locally anisotropic
cosmological models\footnote{%
we shall use also the equivalent terms 'inhomogeneous / nonhomogeneous' and
say that the solutions are (locally) anisotropic if the geometric
constructions are defined for off--diagonal metrics with dependencies both
on time and space coordinates} can be elaborated. The Einstein equations are
very complex systems of nonlinear partial differential equations. If we
perform cosmological approximations for an explicit metric ansatz (for
instance, considering only the dependence on time by averaging on space
coordinates and imposing certain additional spacetime symmetries), we get
some systems of nonlinear ordinary differential equations. Even we may be
able to solve such systems in a general form, we get only a very restricted
subclass of cosmological solutions. This way we "cut" the bulk of nonlinear
gravitational interactions and loose a number of important off--diagonal
solutions.

In our approach, we can integrate the Einstein equations in some very
general forms. Performing approximations for general solutions (and not for
certain coefficients of systems of nonlinear equations), we find new classes
of cosmological metrics which extend the already known families of metrics
for Bianchi universes, Kasner spacetimes etc. We note, that it is not
possible to derive such cosmological metrics working directly with the
Levi--Civita connection and local coordinate frames. The surprising property
of separation of equations\footnote{%
it should be not confused with separation of variables} exists for a more
general type of connections which are also completely (and uniquely, but
with a different geometric meaning) determined by the metric structure. At
the first step, we can construct solutions for generalized connections and
then, the second step, we have to constrain some coefficients of metrics in
order to generate (in our case, cosmological) solutions in Einstein gravity.
Solutions with "un--constrained" off--diagonal metrics also present a
substantial interest for modern cosmology because they can be related to
more general models of string/ brane cosmology etc.

In our further works, we are going to provide an exhaustive study of generic
off--diagonal and locally anisotropic metrics, and related cosmological
models, in Einstein and Lagrange--Finsler theories of gravity that possess a
FRW limit. The purpose of such constructions is to characterize as full as
possible the cosmological spacetimes with anisotropies and generic
off--diagonal nonlinear interactions and thus provide the strongest possible
constraints on exotic cosmologies. In the present paper, however, we focus
on the very specific question of whether general off--diagonal cosmological
solutions can be constructed in Einstein gravity and if such models
necessarily involve de Sitter stadia and possible inflation induced by
nonlinear gravitational interactions.

This paper is organized as follows:

In section \ref{s2}, we present necessary geometric preliminaries on the
nonlinear connection formalism and nonholonomic deformations of metrics,
connections and frames in (pseudo) Riemannian spacetimes. We outline certain
classes of important cosmological solutions which in this work will be
deformed nonholonomically into generic off--diagonal solutions. The Einstein
equations are equivalently formulated for two types of important linear
connections. We use the fact that for the so--called canonical distinguished
connection, the gravitational field equations can be separated with respect
to adapted nonholonomic frames.

In section \ref{s3}, we prove that the Einstein equations can be integrated
in very general forms containing all possible inhomogeneous and locally
anisotropic cosmological solutions. There are analyzed some important
parametrizations and subclasses of such off--diagonal solutions.

Section \ref{s4} is devoted to explicit constructions of generic
off--diagonal cosmological solutions. We derive families of anisotropic
spacetimes containing in certain limits the FRW, Bianchi, Kasner and G\"{o}%
del type configurations. We show that imposing nonintegrable (nonholonomic)
constraints on the nonlinear dynamics of off--diagonal gravitational
interactions we can model various types of anisotropic and de Sitter
solutions.

We summarize and discuss the results in section \ref{s5}.

\section{Nonholonomic Deformations of Cosmological \newline
Solutions}

\label{s2}In this section, we give general features of the geometry
ofnonholonomic deformations and apply this formalism for constructing exact
off--diagonal cosmological solutions. We follow the notations of \cite%
{ijgmmp,vncg,vrflg,vsgg} were details and references can be found.

\subsection{Geometric preliminaries}

Let us consider a (pseudo) Riemannian 4--d manifold $\ \mathbf{V}$ endowed
with a metric $\mathbf{g}=g_{\alpha \beta }(u^{\gamma })du^{\alpha }\otimes
du^{\beta }$ of signature $(+,+,-,+)$ when local coordinates are
parametrized in the form $u^{\alpha }=(x^{i},y^{a}),$ where $%
x^{i}=(x^{1},x^{2})$ and $y^{a}=\left( y^{3}=t,y^{4}=y\right) .$\footnote{%
In our works, we use conventions from \cite{ijgmmp,vrflg} when left up/low
indices are used as labels for spaces and geometric objects. We state that $%
y^3=t$ because such a parametrization will allow us to construct and write
down the formulas for equations and solutions in a "most" simplified form.}
Indices $i,j,k,...=1,2$ and $a,b,c,...=3,4$ are used for a conventional $%
(2+2)$--splitting of dimension and general abstract/coordinate indices when $%
\alpha ,\beta ,\ldots $ run values $1,2,3,4.$

We denote by $\nabla =\{\Gamma _{\ \beta \gamma }^{\alpha }\}$ the
Levi--Civita connection\footnote{%
which is uniquely defined by a given tensor $\mathbf{g}$ to be metric
compatible, $\nabla \mathbf{g}=0,$ and with zero torsion;\ we follow the
conventions established in our previous works \cite{ijgmmp,vrflg}, including
summarizing on ''up-low'' repeating indices if the contrary is not stated}
with coefficients stated with respect to an arbitrary local frame basis $%
e_{\alpha }=(e_{i},e_{a})$ and its dual basis $e^{\beta }=(e^{j},e^{b}).$
Contracting the first and third coefficients of the Riemannian curvature
tensor $\mathcal{R}=\{R_{\ \beta \gamma \delta }^{\alpha }\}$ of $\nabla ,$
we define the Ricci tensor, $\mathcal{R}ic=\{R_{\ \beta \delta }\doteqdot
R_{\ \beta \alpha \delta }^{\alpha }\},$ which (in its turn) can be used for
computing the scalar curvature $R\doteqdot g^{\beta \delta }R_{\ \beta
\delta },$ where $g^{\beta \delta }$ is inverse to $g_{\alpha \beta }.$ The
Einstein equations on $\mathbf{V,}$ for an energy--momentum source of matter
$T_{\alpha \beta },$ are written in the form%
\begin{equation}
R_{\ \beta \delta }-\frac{1}{2}g_{\beta \delta }R=\varkappa T_{\beta \delta
},  \label{einsteq}
\end{equation}%
where $\varkappa =const.$

Our goal is to construct exact solutions of gravitational field equations (%
\ref{einsteq}) parametrized in the form
\begin{eqnarray}
\ ^{\eta }\mathbf{g} &\mathbf{=}&\eta _{i}(x^{k},t)\ ^{\circ }g_{i}(x^{k},t){%
dx^{i}\otimes dx^{i}}+\eta _{a}(x^{k},t)\ ^{\circ }h_{a}(x^{k},t)\mathbf{e}%
^{a}{\otimes }\mathbf{e}^{a},  \label{gsol} \\
\mathbf{e}^{3} &=&dt+\eta _{i}^{3}(x^{k},t)\ ^{\circ }w_{i}(x^{k},t)dx^{i},\
\mathbf{e}^{4}=dy^{4}+\eta _{i}^{4}(x^{k},t)\ ^{\circ }n_{i}(x^{k},t)dx^{i},
\notag
\end{eqnarray}
for certain classes of coefficients (functions) to be defined below. In
brief, we shall write such metrics as
\begin{equation}
\ \mathbf{g}\mathbf{=}g_{ij}{dx^{i}\otimes dx^{j}}%
+h_{ab}(dy^{a}+N_{k}^{a}dx^{k}){\otimes }(dy^{b}+N_{k}^{b}dx^{k}),
\label{dm}
\end{equation}%
where, for (\ref{gsol}), $g_{ij}=diag[g_{i}=\eta _{i}\ ^{\circ }g_{i}]$ and $%
h_{ab}=diag[h_{a}=\eta _{a}\ ^{\circ }h_{a}]$ and $N_{k}^{3}=w_{i}=\eta
_{i}^{3}\ ^{\circ }w_{i}$ and $N_{k}^{4}=n_{i}=\eta _{i}^{4}\ ^{\circ
}n_{i}. $ The gravitational 'polarizations' $\eta _{\alpha }$ and $\eta
_{i}^{a}$ determine nonholonomic deformations of metrics, $\ ^{\circ }%
\mathbf{g}\mathbf{=}[\ ^{\circ }g_{i},\ ^{\circ }h_{a},\ ^{\circ
}N_{k}^{a}]\rightarrow \ ^{\eta }\mathbf{g}\mathbf{=}[\
g_{i},h_{a},N_{k}^{a}].$ Such transforms (with deformations of the frame,
metric, connections and other fundamental geometric structures) are more
general than those considered for the Cartan's moving frame method, when the
geometric objects are re--defined equivalently with respect to necessary
systems of reference.

Any set of coefficients $N_{k}^{a}$ in (\ref{dm}) state on $\mathbf{V}$ some
$N$--adapted frame, $\mathbf{e}_{\alpha },$ and dual frame, $\mathbf{e}_{\
}^{\beta },$ structures (i.e. N--elongated partial derivatives,
respectively, differentials)
\begin{eqnarray}
\mathbf{e}_{\alpha } &\doteqdot &\left( \mathbf{e}_{i}=\partial
_{i}-N_{i}^{a}\partial _{a},e_{b}=\partial _{b}=\frac{\partial }{\partial
y^{b}}\right) ,  \label{ddr} \\
\mathbf{e}_{\ }^{\beta } &\doteqdot &\left( e^{i}=dx^{i},\mathbf{e}%
^{a}=dy^{a}+N_{i}^{a}dx^{i}\right) .  \label{ddf}
\end{eqnarray}%
Such local bases satisfy some nonholonomic relations
\begin{equation}
\left[ \mathbf{e}_{\alpha },\mathbf{e}_{\beta }\right] =\mathbf{e}_{\alpha }%
\mathbf{e}_{\beta }-\mathbf{e}_{\beta }\mathbf{e}_{\alpha }=\mathbf{w}_{\
\alpha \beta }^{\gamma }\left( u\right) \mathbf{e}_{\gamma },  \label{anhr}
\end{equation}%
with nontrivial anholonomy coefficients $\mathbf{w}_{\beta \gamma }^{\alpha
}\left( u\right),$
\begin{equation}
\mathbf{w}_{~ji}^{a}=-\mathbf{w}_{~ij}^{a}=\Omega _{ij}^{a}=\mathbf{e}%
_{j}N_{i}^{a}-\mathbf{e}_{i}N_{j}^{a},\ \mathbf{w}_{~ia}^{b}=-\mathbf{w}%
_{~ai}^{b}=\partial _{a}N_{i}^{b}.  \label{anhc}
\end{equation}

We can fix on $\mathbf{V}$ such systems of N--elongated frames when the sets
of coefficients $\mathbf{N}=\{N_{k}^{a}\}$ define a Whitney splitting (in
general, non--integrable) of tangent space $T\mathbf{V}$ to $\mathbf{V,}$%
\begin{equation}
T\mathbf{V}=h\mathbf{V\oplus }v\mathbf{V}  \label{whitney}
\end{equation}%
into conventional horizontal (h) and vertical (v) subspaces, respectively, $h%
\mathbf{V}$ and $v\mathbf{V.}$ Such a geometric object defines a nonlinear
connection (N--connection) structure.

For any metric $\mathbf{g}$ (\ref{dm}) on a spacetime $\mathbf{V,}$ there is
an infinite number of metric compatible linear connections $D,$ satisfying
the conditions $D\mathbf{g}=0,$ and completely defined by $\mathbf{g.}$ A
subclass of such linear connections can be adapted to a chosen N--connection
structure $\mathbf{N,}$ when the splitting (\ref{whitney}) is preserved
under parallelism, and called distinguished connections (in brief,
d--connections). A general d--connection is denoted by a boldface symbol $%
\mathbf{D}=(hD,vD)$ distinguished into, respectively, h- and v--covariant
derivatives, $hD$ and $vD.$ To construct exact solutions in gravity theories
is convenient to work with the so--called canonical d--connection, $\widehat{%
\mathbf{D}}=\{\widehat{\mathbf{\Gamma }}_{\ \alpha \beta }^{\gamma }\},$
which with respect to N--adapted bases (\ref{ddr}) and (\ref{ddf}) is given
by coefficients $\widehat{\mathbf{\Gamma }}_{\ \alpha \beta }^{\gamma
}=\left( \widehat{L}_{jk}^{i},\widehat{L}_{bk}^{a},\widehat{C}_{jc}^{i},%
\widehat{C}_{bc}^{a}\right) ,$ for $h\widehat{D}=\{\widehat{L}_{jk}^{i},%
\widehat{L}_{bk}^{a}\}$ and $v\widehat{D}=\{\widehat{C}_{jc}^{i},\widehat{C}%
_{bc}^{a}\},$ where
\begin{eqnarray}
\widehat{L}_{jk}^{i} &=&\frac{1}{2}g^{ir}\left(
e_{k}g_{jr}+e_{j}g_{kr}-e_{r}g_{jk}\right),  \label{candcon} \\
\widehat{L}_{bk}^{a} &=&e_{b}(N_{k}^{a})+\frac{1}{2}h^{ac}\left(
e_{k}h_{bc}-h_{dc}\ e_{b}N_{k}^{d}-h_{db}\ e_{c}N_{k}^{d}\right) ,  \notag \\
\widehat{C}_{jc}^{i} &=&\frac{1}{2}g^{ik}e_{c}g_{jk},\ \widehat{C}_{bc}^{a}=%
\frac{1}{2}h^{ad}\left( e_{c}h_{bd}+e_{c}h_{cd}-e_{d}h_{bc}\right) .  \notag
\end{eqnarray}%
This canonical d--connection $\widehat{\mathbf{D}}$ and its torsion $%
\mathcal{T}=\{\widehat{\mathbf{T}}_{\ \alpha \beta }^{\gamma }\equiv
\widehat{\mathbf{\Gamma }}_{\ \alpha \beta }^{\gamma }-\widehat{\mathbf{%
\Gamma }}_{\ \beta \alpha }^{\gamma };\newline
\widehat{T}_{\ jk}^{i}, \widehat{T}_{\ ja}^{i},\widehat{T}_{\ ji}^{a},%
\widehat{T}_{\ bi}^{a},\widehat{T}_{\ bc}^{a}\},$ where the nontrivial
coefficients
\begin{eqnarray}
\widehat{T}_{\ jk}^{i} &=&\widehat{L}_{jk}^{i}-\widehat{L}_{kj}^{i},\widehat{%
T}_{\ ja}^{i}=\widehat{C}_{jb}^{i},\widehat{T}_{\ ji}^{a}=-\Omega _{\
ji}^{a},  \label{dtors} \\
\widehat{T}_{aj}^{c} &=&\widehat{L}_{aj}^{c}-e_{a}(N_{j}^{c}),\widehat{T}_{\
bc}^{a}=\ \widehat{C}_{bc}^{a}-\ \widehat{C}_{cb}^{a},  \notag
\end{eqnarray}%
are completely defined by the coefficients of metric $\mathbf{g}$ (\ref{dm})
following the conditions that $\widehat{\mathbf{D}}\mathbf{g=}0$ and the
''pure'' horizontal and vertical torsion coefficients are zero, i. e. $%
\widehat{T}_{\ jk}^{i}=0$ and $\widehat{T}_{\ bc}^{a}=0.$

Any geometric construction for the canonical d--connection $\widehat{\mathbf{%
D}}$ can be re--defined equivalently into a similar one with the
Levi--Civita connection following formula
\begin{equation}
\Gamma _{\ \alpha \beta }^{\gamma }=\widehat{\mathbf{\Gamma }}_{\ \alpha
\beta }^{\gamma }+Z_{\ \alpha \beta }^{\gamma },  \label{deflc}
\end{equation}%
where the distortion tensor $Z_{\ \alpha \beta }^{\gamma }$ is constructed
in a unique form from the coefficients of a metric $\mathbf{g}_{\alpha \beta
},$
\begin{eqnarray}
\ Z_{jk}^{a} &=&-\widehat{C}_{jb}^{i}g_{ik}h^{ab}-\frac{1}{2}\Omega
_{jk}^{a},~Z_{bk}^{i}=\frac{1}{2}\Omega _{jk}^{c}h_{cb}g^{ji}-\Xi _{jk}^{ih}~%
\widehat{C}_{hb}^{j},  \notag \\
Z_{bk}^{a} &=&~^{+}\Xi _{cd}^{ab}~\widehat{T}_{kb}^{c},\ Z_{kb}^{i}=\frac{1}{%
2}\Omega _{jk}^{a}h_{cb}g^{ji}+\Xi _{jk}^{ih}~\widehat{C}_{hb}^{j},\
Z_{jk}^{i}=0,  \label{deft} \\
\ Z_{jb}^{a} &=&-~^{-}\Xi _{cb}^{ad}~\widehat{T}_{jd}^{c},\ Z_{bc}^{a}=0,\
Z_{ab}^{i}=-\frac{g^{ij}}{2}\left[ \widehat{T}_{ja}^{c}h_{cb}+\widehat{T}%
_{jb}^{c}h_{ca}\right] ,  \notag
\end{eqnarray}%
for $\ \Xi _{jk}^{ih}=\frac{1}{2}(\delta _{j}^{i}\delta
_{k}^{h}-g_{jk}g^{ih})$ and $~^{\pm }\Xi _{cd}^{ab}=\frac{1}{2}(\delta
_{c}^{a}\delta _{d}^{b}+h_{cd}h^{ab}).$

\subsection{Limits to known cosmological solutions}

\label{sskesol}In this work, we shall construct new classes of cosmological
solutions\footnote{%
in general, such spacetimes inhomogeneous and anisotropic} with metrics $\
^{\eta }\mathbf{g},$ i.e. 'target' metrics, possessing certain limits, for $%
\eta _{\alpha },$ $\eta _{i}^{a}\rightarrow 1,$ or $\eta _{i}^{a}\rightarrow
0,$ to $\ ^{\circ }\mathbf{g}$ (a 'prime' metric) which is a conformal, or
frame/coordinate, transform of a well known FRW, Bianchi, Kasner, or another
type solution, see reviews of results in \cite{calog,sungcoles1,mukh,weinb}.

\subsubsection{FRW metrics}

The FRW cosmological solution can be written in the form%
\begin{equation}
\ \ _{F}\mathbf{g}\mathbf{=}a^{2}(t)\left( \frac{dr{\otimes }dr}{1-\kappa
r^{2}}+r^{2}d\theta {\otimes }d\theta \right) -dt{\otimes }%
dt+a^{2}(t)r^{2}\sin ^{2}\theta d\varphi {\otimes }d\varphi ,  \label{frw}
\end{equation}%
with $\kappa =\pm 1,0,$ when the coordinates and coefficients of metric are
paramet\-rized, respectively, in the form $x^{1}=r,x^{2}=\theta
,y^{3}=t,y^{4}=\varphi $ (for spherical coordinates) and $\ \ \
_{F}g_{1}=a^{2}/(1-\kappa r^{2}),$ $\ \ \ _{F}g_{2}=a^{2}r^{2}/(1-\kappa
r^{2}),\ \ \ _{F}h_{3}=-1,\ \ \ _{F}h_{4}=a^{2}(t)r^{2}\sin ^{2}\theta $ and
$\ \ \ _{F}N_{i}^{a}=0.$\footnote{%
Instead of FRW as a 'prime' metric $\ ^{\circ }\mathbf{g}$, we can consider
any Biachi, Kasner etc cosmological solutions outlined in \cite%
{sungcoles1,pontzen,odintsov,sotr,capozz} and references therein.} This
metric is an exact solution of equations (\ref{einsteq}) with a perfect
fluid stress--energy tensor,%
\begin{equation}
T_{\ \ \beta }^{\alpha }=diag[-p,-p,\rho ,-p],  \label{fluid}
\end{equation}%
where $\rho $ and $p$ are the proper energy density and pressure in the
fluid rest frame. The Einstein equations for ansatz (\ref{frw}) take the
form of two coupled nonlinear ordinary differential equations (also called
the Friedmann equations)%
\begin{equation}
H^{2}\equiv \left( \frac{a^{\ast }}{a}\right) ^{2}=\frac{1}{3}\rho -\frac{%
\kappa }{a^{2}}  \label{fr1}
\end{equation}%
and
\begin{equation}
H^{\ast }+H^{2}=\frac{a^{\ast \ast }}{a}=-\frac{1}{6}(\rho +3p),  \label{fr2}
\end{equation}%
where the strong energy conditions for matter, $\rho +3p\geq 0,$ must be
satisfied for an expanding universe.\footnote{%
For our purposes, and following our former works on geometric methods in
gravity and exact solutions \cite{vgsolhd,vncg,ijgmmp}, it is convenient to
introduce a system of notations which is different form those used in
standard books on cosmology (see, for instance, \cite%
{mukh,weinb,odintsov,sotr,capozz,wmap,lambda,planck}, were readers may
consult details and references on modern cosmology).} The Hubble parameter $%
H\equiv \frac{a^{\ast }}{a}$ has the unit of inverse time and is positive
(negative) for an expanding (collapsing) universe. The equations (\ref{fr1})
and (\ref{fr2}) may be combined (which is also related to the condition $%
\nabla _{\alpha }T_{\ \ \beta }^{\alpha }=0)$%
\begin{equation*}
\rho ^{\ast }+3H(\rho +p)=0.
\end{equation*}

For simplicity, we can consider $\kappa =0$ with
\begin{equation}
\ \ \ \ _{F}\mathbf{g}\mathbf{=}a^{2}(t)\left( dx{\otimes }dx+dz{\otimes }%
dz\right) -dt{\otimes }dt+a^{2}(t)dy{\otimes }dy,  \label{frw1}
\end{equation}%
for Cartezian coordinates and coefficients of metric paramet\-rized,
respectively, in the form $x^{1}=x,x^{2}=z,y^{3}=t,y^{4}=y$ and $\ ^{\circ
}g_{1}=$ $\ \ \ _{F}g_{2}=a^{2},\ \ \ _{F}h_{3}=-1$ and $\ \ \
_{F}h_{4}=a^{2}$ (in this case, the nontrivial coefficients of metric depend
only on time like coordinate, $t,$ but not on space like ones).

\subsubsection{Bianchi type metrics}

All possible spatially homogeneous but anisotropic relativistic cosmological
models are arranged following the Bianchi classification corresponding to
symmetry properties of their spatial hypersurfaces \cite%
{grish,ellis,sungcoles1}. Such cosmologocal metrics can be parametrized by
orthonormal tetrad (vierbein) bases $e_{\alpha ^{\prime \prime }}=e_{\
\alpha ^{\prime \prime }}^{\underline{\alpha }}\partial /\partial u^{%
\underline{\alpha }},$ when
\begin{equation}
\ \ _{B}\mathbf{g}_{\alpha ^{\prime \prime }\beta ^{\prime \prime }}=\ \
_{B}e_{\ \alpha ^{\prime \prime }}^{\underline{\alpha }}\ \ _{B}e_{\ \beta
^{\prime \prime }}^{\underline{\beta }}\ \ _{B}g_{\underline{\alpha }%
\underline{\beta }}=diag[1,1,-1,1]  \label{bianchim}
\end{equation}%
and
\begin{equation*}
\left[ \ \ _{B}e_{\alpha ^{\prime \prime }},\ \ _{B}e_{\beta ^{\prime \prime
}}\right] =\ _{B}w_{\ \alpha ^{\prime \prime }\beta ^{\prime \prime
}}^{\gamma ^{\prime \prime }}\left( t\right) \ \ _{B}e_{\gamma ^{\prime
\prime }},
\end{equation*}%
with time dependent 'structure constants'
\begin{equation}
\ \ \ _{B}w_{\ \alpha ^{\prime \prime }\beta ^{\prime \prime }}^{\gamma
^{\prime \prime }}\left( t\right) =\ \epsilon _{\ \alpha ^{\prime \prime
}\beta ^{\prime \prime }\tau ^{\prime \prime }}n^{\tau ^{\prime \prime
}\gamma ^{\prime \prime }}\left( t\right) +\delta _{\beta ^{\prime \prime
}}^{\gamma ^{\prime \prime }}b_{\alpha ^{\prime \prime }}\left( t\right)
-\delta _{\alpha ^{\prime \prime }}^{\gamma ^{\prime \prime }}b_{\beta
^{\prime \prime }}\left( t\right)  \label{bstrc}
\end{equation}%
(determined by some diagonal tensor, $n^{\tau ^{\prime \prime }\gamma
^{\prime \prime }},$ and vector, $b_{\alpha ^{\prime \prime }},$ fields)
used for the mentioned classification. Depending on parametrization of such
tensor and vector objects, there are constructed the so--called Bianchi
universes which are closed, or not, to the homogeneous and isotropic FRW
case. With nontrivial limits to observable cosmology, there are the so
--called Bianchi $I,V,VII_{0},VII_{h}$ and $IX$ cosmologies.

\subsubsection{Kasner type metrics}

For instance, in four dimensional gravity, such a metric is written in the
form%
\begin{equation}
\ \ \ \ _{K}\mathbf{g}\mathbf{=}t^{2p_{1}}dx{\otimes }dx+t^{2p_{3}}dz{%
\otimes }dz-dt{\otimes }dt+t^{2p_{2}}dy{\otimes }dy,  \label{kasner}
\end{equation}%
with $\ \ \ _{K}g_{1}=t^{2p_{1}},\ _{K}g_{2}=t^{2p_{3}},\ \ _{K}h_{3}=-1,\
_{K}h_{4}=t^{2p_{2}},\ _{K}N_{i}^{a}=0,$ see details (including modern brane
generalizations) and references in \cite{frolov,roberts,steinh}. The
constants $p_{1},p_{2},p_{3}$ define solutions of the Einstein equations if
there are satisfied the conditions%
\begin{equation}
2\ ^{3}P=\ ^{2}P-\ ^{1}P,  \label{kasner1}
\end{equation}%
for $\left( \ ^{1}P\right) ^{2}=\left( p_{1}\right) ^{2}+\left( p_{2}\right)
^{2}+\left( p_{3}\right) ^{2},\ ^{2}P=p_{1}+p_{2}+p_{3},\
^{3}P=p_{1}p_{2}+p_{2}p_{3}+p_{1}p_{3}.$ Following the anholonomic
deformation method, we shall generalize such solutions to generic
off--diagonal cosmological configurations, see section \ref{ofanissol}.

\subsubsection{G\"{o}del model}

The theoretical model for the study of rotating cosmology is determined by
the G\"{o}del solution \cite{godel}
\begin{equation}
\ \ \ _{G}\mathbf{g}\mathbf{=}\ \ \ \ _{G}a^{2}\left[ dx{\otimes }dx+\frac{%
e^{2x}}{2}dz{\otimes }dz-(dt-e^{x}dz){\otimes }(dt-e^{x}dz)+dy{\otimes }dy%
\right] ,  \label{godele}
\end{equation}%
when $\ \ \ _{G}g_{1}=\ \ \ _{G}a^{2},$ $\ \ \ _{G}g_{2}=\frac{e^{2x}}{2}\ \
\ _{G}a^{2},\ \ \ _{G}h_{3}=-\ \ \ _{G}a^{2},\ \ \ _{G}h_{4}=\ \ \
_{G}a^{2}, $ $\ \ \ _{G}N_{i}^{a}=\ \ \ _{G}w_{i}=-e^{x},$ $\ \ \
_{G}N_{i}^{a}=0.$ The matter is described as a dust with the energy density $%
\ \ \ _{G}\varepsilon $ and there is a nontrivial negative (with opposite
sign to that introduced by Einstein) cosmological constant $\ \ \
_{G}\lambda $ determining the angular velocity $\ \ \ _{G}\omega $ of the
cosmic rotation, $\ \ \ _{G}\omega ^{2}=1/2\ \ \ _{G}a^{2}=4\pi G\ \ \
_{G}\varepsilon =-\ \ \ _{G}\lambda ,$ with $G$ as Newton's gravitational
constant. The parametrizations of coordinates and coefficients of metrics
are different from that considered in the former works, see a comprehencive
review of results and references in \cite{obukh1,obukhov,barrow}. In section %
\ref{ofanissol}, we shall provide an off--diagonal generalization defining
rotating universes with polarized cosmological constants and
nonholonomic/constrained rotations in nontrivial backgrounds.

\subsection{The Einstein equations for connections $\widehat{\mathbf{D}}$
and $\protect\nabla $}

The Einstein equations (\ref{einsteq}) for a metric $\mathbf{g}_{\beta
\delta }$ can be rewritten equivalently using the canonical d--connection $%
\widehat{\mathbf{D}},$%
\begin{eqnarray}
&&\widehat{\mathbf{R}}_{\ \beta \delta }-\frac{1}{2}\mathbf{g}_{\beta \delta
}\ ^{s}R=\mathbf{\Upsilon }_{\beta \delta },  \label{cdeinst} \\
&&\widehat{L}_{aj}^{c}=e_{a}(N_{j}^{c}),\ \widehat{C}_{jb}^{i}=0,\ \Omega
_{\ ji}^{a}=0,  \label{lcconstr}
\end{eqnarray}%
where $\widehat{\mathbf{R}}_{\ \beta \delta }$ is the Ricci tensor for $%
\widehat{\mathbf{\Gamma }}_{\ \alpha \beta }^{\gamma },\ ^{s}R=\mathbf{g}%
^{\beta \delta }\widehat{\mathbf{R}}_{\ \beta \delta }$ and $\mathbf{%
\Upsilon }_{\beta \delta }$ is such way constructed that $\mathbf{\Upsilon }%
_{\beta \delta }\rightarrow \varkappa T_{\beta \delta }$ \ for $\widehat{%
\mathbf{D}}\rightarrow \nabla .$ We emphasize here that if the constraints (%
\ref{lcconstr}) are satisfied the tensors $\widehat{\mathbf{T}}_{\ \alpha
\beta }^{\gamma }$(\ref{dtors}) and $Z_{\ \alpha \beta }^{\gamma }$(\ref%
{deft}) are zero. This states that $\widehat{\mathbf{\Gamma }}_{\ \alpha
\beta}^{\gamma }=\Gamma _{\ \alpha \beta }^{\gamma },$  with respect to
N--adapted frames (\ref{ddr}) and (\ref{ddf}), see (\ref{deflc}), even $%
\widehat{\mathbf{D}}\neq \nabla .$

In a series of our works \cite{vncg,ijgmmp,vsgg,vrflg,vgsolhd}, we provided
detailed proofs that for constructing exact solutions with generic
off--diagonal metrics it is more convenient to work with the canonical
d--connection $\widehat{\mathbf{D}}$ than with the Levi--Civita connection $%
\nabla ;$ the last one is considered to be the standard one for general
relativity. The surprising thing is that the ''nonholonomic'' gravitational
field equations (\ref{cdeinst}) split in such a form, with respect to
N--adapted frames (\ref{ddr}) and (\ref{ddf}), that the resulting system of
partial differential equations (see below, for instance, (\ref{eq1}) -- (\ref%
{eq4})) can be solved in very general forms. In order to generate exact
solutions for $\nabla ,$ we have to impose additional constraints (\ref%
{lcconstr}) on coefficients of metric $\mathbf{g}$ (\ref{dm}) (for instance,
on $\ ^{\eta }\mathbf{g}$ \ (\ref{gsol}) if we wont to generate new classes
of cosmological solutions).\footnote{%
We note that it is not possible to solve the Einstein equations in general
form working directly with $\nabla $ because this way we do not get a
separation of nonlinear partial differential equations. Our idea is to use a
more general connection (also defined completely by the same metric), when
the system of nonlinear equations can be integrated in some general forms,
and than to constrain the solutions to generate metrics for the general
relativity theory.}

For an ansatz of type (\ref{gsol}), the Einstein equations (\ref{cdeinst})
for $\widehat{\mathbf{D}}$ with a general source of type\footnote{%
such parametrizations of energy--momentum tensors are possible by
corresponding nonholonomic frame and/or coordinate frame transform for very
general matter sources, including some important cases with cosmological
constants and various models of locally anisotropic fluid/scalar field/
spinor/ gauge fields interactions on curved spaces}
\begin{equation}
\Upsilon _{\ \ \beta }^{\alpha }=diag[\Upsilon _{\gamma };\Upsilon
_{1}=\Upsilon _{2}=\Upsilon _{2}(x^{k},t);\Upsilon _{3}=\Upsilon
_{4}=\Upsilon _{4}(x^{k})]  \label{source}
\end{equation}
transform into a system of nonlinear partial differential equations with
separation of equations for h-- and v--components of metric and
N--connection coefficients,
\begin{eqnarray}
\widehat{R}_{1}^{1} &=&\widehat{R}_{2}^{2}  \label{eq1} \\
&=&-\frac{1}{2g_{1}g_{2}}\left[ g_{2}^{\bullet \bullet }-\frac{%
g_{1}^{\bullet }g_{2}^{\bullet }}{2g_{1}}-\frac{\left( g_{2}^{\bullet
}\right) ^{2}}{2g_{2}}+g_{1}^{\prime \prime }-\frac{g_{1}^{l}g_{2}^{l}}{%
2g_{2}}-\frac{\left( g_{1}^{l}\right) ^{2}}{2g_{1}}\right] =-\Upsilon
_{4}(x^{k}),  \notag \\
\widehat{R}_{3}^{3} &=&\widehat{R}_{4}^{4}=-\frac{1}{2h_{3}h_{4}}\left[
h_{4}^{\ast \ast }-\frac{\left( h_{4}^{\ast }\right) ^{2}}{2h_{4}}-\frac{%
h_{3}^{\ast }h_{4}^{\ast }}{2h_{3}}\right] =-\Upsilon _{2}(x^{k},t),
\label{eq2}
\end{eqnarray}
\begin{eqnarray}
\widehat{R}_{3k} &=&\frac{w_{k}}{2h_{4}}\left[ h_{4}^{\ast \ast }-\frac{%
\left( h_{4}^{\ast }\right) ^{2}}{2h_{4}}-\frac{h_{3}^{\ast }h_{4}^{\ast }}{%
2h_{3}}\right]  \label{eq3} \\
&&+\frac{h_{4}^{\ast }}{4h_{4}}\left( \frac{\partial _{k}h_{3}}{h_{3}}+\frac{%
\partial _{k}h_{4}}{h_{4}}\right) -\frac{\partial _{k}h_{4}^{\ast }}{2h_{4}}%
=0,  \notag \\
\widehat{R}_{4k} &=&\frac{h_{4}}{2h_{3}}n_{k}^{\ast \ast }+\left( \frac{h_{4}%
}{h_{3}}h_{3}^{\ast }-\frac{3}{2}h_{4}^{\ast }\right) \frac{n_{k}^{\ast }}{%
2h_{3}}=0,  \label{eq4}
\end{eqnarray}%
In brief, we wrote the partial derivatives in the form $a^{\bullet
}=\partial a/\partial x^{1},$\ $a^{\prime }=\partial a/\partial x^{2},$\ $%
a^{\ast }=\partial a/\partial t.$ The ansatz (\ref{gsol}) does not depend on
variable $y^{4}$ (that why we do not have terms with $\partial /\partial
y^{4}$).

The above system of equations can be solved in very general forms for
arbitrary dimensions and signatures as we proved in Refs. \cite%
{vncg,ijgmmp,vsgg,vrflg,vgsolhd} (see also next section). In this work, we
analyze ''cosmological configurations'' for $\widehat{\mathbf{D}}$ when $%
y^{3}=t$ for generic off--diagonal metrics of type (\ref{gsol}). New classes
of cosmological conditions in general relativity, with the Levi--Civita
connection $\nabla ,$ will be extracted by imposing additional constraints
\begin{equation}
w_{i}^{\ast }=\mathbf{e}_{i}\ln |h_{4}|,\mathbf{e}_{k}w_{i}=\mathbf{e}%
_{i}w_{k},\ n_{i}^{\ast }=0,\ \partial _{i}n_{k}=\partial _{k}n_{i}
\label{lcconstr1}
\end{equation}
satisfying the conditions (\ref{lcconstr}).

\section{General Cosmological Off--Diagonal Solutions}

\label{s3} In this section, we construct in explicit form and analyze the
properties of possible classes of solutions depending on time like variable $%
t$ and on "horizontal" spacelike coordinates $x^{i}$ of gravitational field
equations (\ref{eq1})--(\ref{eq4}), for the canonical d--connection, and of
constraints (\ref{lcconstr1}) selecting Levi--Civita configurations.

\subsection{Type 1: Solutions with $h_{3,4}^{\ast }\neq 0$ and $\Upsilon
_{2,4}\neq 0$}

Such metrics are defined by a metric ansatz
\begin{eqnarray}
\ ^{\eta }\mathbf{g} &\mathbf{=}&e^{\psi (x^{k})}{dx^{i}\otimes dx^{i}}%
+h_{3}(x^{k},t)\mathbf{e}^{3}{\otimes }\mathbf{e}^{3}+h_{4}(x^{k},t)\mathbf{e%
}^{4}{\otimes }\mathbf{e}^{4},  \label{genans} \\
\mathbf{e}^{3} &=&dt+w_{i}(x^{k},t)dx^{i},\mathbf{e}%
^{4}=dy^{4}+n_{i}(x^{k},t)dx^{i}  \notag
\end{eqnarray}%
with the coefficients being solutions of the system\footnote{%
it is an equivalent of equations (\ref{eq1})--(\ref{eq4}) for $h_{3,4}^{\ast
}\neq 0$}
\begin{eqnarray}
\ddot{\psi}+\psi ^{\prime \prime } &=&2\Upsilon _{4}(x^{k}),  \label{4ep1a}
\\
h_{4}^{\ast } &=&2h_{3}h_{4}\Upsilon _{2}(x^{i},t)/\phi ^{\ast },
\label{4ep2a}
\end{eqnarray}
\begin{eqnarray}
\beta w_{i}+\alpha _{i} &=&0,  \label{4ep3a} \\
n_{i}^{\ast \ast }+\gamma n_{i}^{\ast } &=&0,  \label{4ep4a}
\end{eqnarray}%
where%
\begin{equation}
~\phi =\ln |\frac{h_{4}^{\ast }}{\sqrt{|h_{3}h_{4}|}}|,\ \alpha
_{i}=h_{4}^{\ast }\partial _{i}\phi ,\ \beta =h_{4}^{\ast }\ \phi ^{\ast },\
\gamma =\left( \ln |h_{4}|^{3/2}/|h_{3}|\right) ^{\ast }.  \label{auxphi}
\end{equation}%
For $h_{4}^{\ast }\neq 0;\Upsilon _{2}\neq 0,$ we get $\phi ^{\ast }\neq 0.$
Prescribing any nonconstant $\phi =\phi (x^{i},t)$ as a generating function,
we can construct exact solutions of (\ref{4ep1a})--(\ref{4ep4a}) solving
respectively the two dimensional Laplace equation, for $g_{1}=g_{2}=e^{\psi
(x^{k})};$ integrating on $t,$ in order to determine $h_{3},$ $h_{4}$ and $%
n_{i},$ and solving algebraic equations, for $w_{i}.$ We obtain (computing
consequently for a chosen $\phi (x^{k},t)$)
\begin{eqnarray}
g_{1} &=&g_{2}=e^{\psi (x^{k})},h_{3}=\pm \ \frac{|\phi ^{\ast }(x^{i},t)|}{%
\Upsilon _{2}},\   \label{gsol1} \\
h_{4} &=&\ ^{0}h_{4}(x^{k})\pm \ 2\int \frac{(\exp [2\ \phi
(x^{k},t)])^{\ast }}{\Upsilon _{2}}dt,\   \notag \\
w_{i} &=&-\partial _{i}\phi /\phi ^{\ast },\ n_{i}=\ ^{1}n_{k}\left(
x^{i}\right) +\ ^{2}n_{k}\left( x^{i}\right) \int [h_{3}/(\sqrt{|h_{4}|}%
)^{3}]dt,  \notag
\end{eqnarray}%
where $\ ^{0}h_{4}(x^{k}),\ ^{1}n_{k}\left( x^{i}\right) $ and $\
^{2}n_{k}\left( x^{i}\right) $ are integration functions. We have to fix a
corresponding sign $\pm $ in order to generate a necessary local signature
of type $(++-+)$ for some chosen $\phi ,\Upsilon _{2}$ and $\Upsilon _{4}.$

Here we note that the general off--diagonal solutions (\ref{frw}) include \
as particular cases the solutions for a nontrivial cosmological constant $%
\Upsilon _{i}=\lambda ,$ or nonholonomic configurations with polarizations
of such constants, $\lambda \rightarrow \ ^{h}\lambda (x^{k})=\Upsilon
_{4}(x^{k})$ and $\lambda \rightarrow \ ^{v}\lambda (x^{k},t)=\Upsilon
_{2}(x^{k},t).$

In order to construct exact solutions for the Levi--Civita connection, we
have to constrain the coefficients (\ref{gsol1}) of metric (\ref{genans}) to
satisfy the conditions (\ref{lcconstr1}). Such constraints restrict the
class of generating and integration functions. For instance, we can chose
that $\ ^{2}n_{k}\left( x^{i}\right) =0$ and $\ ^{1}n_{k}\left( x^{i}\right)
$ are any functions satisfying the conditions $\ \partial _{i}\
^{1}n_{k}=\partial _{k}\ ^{1}n_{i}.$ The constraints for $\phi (x^{k},t)$
follow from constraints on N--connection coefficients $w_{i}=-\partial
_{i}\phi /\phi ^{\ast },$
\begin{eqnarray}
\left( w_{i}[\phi ]\right) ^{\ast }+w_{i}[\phi ]\left( h_{4}[\phi ]\right)
^{\ast }+\partial _{i}h_{4}[\phi ]=0, &&  \notag \\
\partial _{i}\ w_{k}[\phi ]=\partial _{k}\ w_{i}[\phi ], &&  \label{auxc1}
\end{eqnarray}%
where, for instance, we denoted by $h_{4}[\phi ]$ the functional dependence
on $\phi .$ Such conditions are always satisfied for cosmological solutions
with $\phi =\phi (t)$ or if $\phi =const$ (in the last case $w_{i}(x^{k},t)$
can be any functions as follows from (\ref{4ep3a}) with zero $\beta $ and $%
\alpha _{i},$ see (\ref{auxphi})).

\subsection{Important special cases}

We can construct such solutions for certain special parametrizations of
coefficients for ansatz (\ref{gsol1}) subjected to the condition to be a
solution of equations (\ref{4ep1a})--(\ref{4ep4a}).

\subsubsection{Type 2: Solutions with $h_{4}^{\ast }=0$}

The equation (\ref{eq2}) can be solved for such a case, $h_{4}^{\ast }=0,$
only if $\Upsilon _{2}=0.$ We can consider any functions $w_{i}(x^{k},t)$ as
solutions of (\ref{eq3}), and its equivalent (\ref{4ep3a}), because the
coefficients $\beta $ and $\alpha _{i},$ see (\ref{auxphi}), are zero. To
find nontrivial values of $n_{i}$ we can integrate (\ref{4ep4a}) for $%
h_{4}^{\ast }=0$ and any given $h_{3}$ which results in $n_{i}=\
^{1}n_{k}\left( x^{i}\right) +\ ^{2}n_{k}\left( x^{i}\right) \int h_{3}dt.$
We can consider any $g_{1}=g_{2}=e^{\psi (x^{k})},$ with $\psi (x^{k})$
determined by (\ref{4ep1a}) for a given $\Upsilon _{4}(x^{k}).$

Summarizing the results, we conclude that this class of solutions is defined
by an ansatz
\begin{eqnarray}
\ ^{\eta }\mathbf{g} &\mathbf{=}&e^{\psi (x^{k})}{dx^{i}\otimes dx^{i}}%
+h_{3}(x^{k},t)\mathbf{e}^{3}{\otimes }\mathbf{e}^{3}+\ ^{0}h_{4}(x^{k})%
\mathbf{e}^{4}{\otimes }\mathbf{e}^{4},  \label{genans1} \\
\mathbf{e}^{3} &=&dt+w_{i}(x^{k},t)dx^{i},\ \mathbf{e}^{4}=dy^{4}+\left[ \
^{1}n_{k}\left( x^{i}\right) +\ ^{2}n_{k}\left( x^{i}\right) \int h_{3}dt%
\right] dx^{i},  \notag
\end{eqnarray}%
for arbitrary generating functions $h_{3}(x^{k},t),w_{i}(x^{k},t),\
^{0}h_{4}(x^{k})$ and integration functions $\ ^{1}n_{k}\left( x^{i}\right) $
and $\ ^{2}n_{k}\left( x^{i}\right) .$

The conditions (\ref{lcconstr1}) selecting from (\ref{genans1}) \ a subclass
of solutions for the Levi--Civita connection transform into the equations
\begin{eqnarray*}
\ ^{2}n_{k}\left( x^{i}\right) =0\ &\mbox{ and }& \partial _{i}\
^{1}n_{k}=\partial _{k}\ ^{1}n_{i}, \\
w_{i}^{\ast }+\partial _{i}\ ^{0}h_{4}=0 &\mbox{ and }& \partial _{i}\
w_{k}=\partial _{k}\ w_{i},
\end{eqnarray*}%
for any such $w_{i}(x^{k},t)$ and $\ ^{0}h_{4}(x^{k}).$ This class of
constraints do not involve the generating function $h_{3}(x^{k},t).$

\subsubsection{Type 3: Solutions with $h_{3}^{\ast }=0$ and $h_{4}^{\ast
}\neq 0$}

Such metrics are defined by ansatz of type
\begin{eqnarray}
\ ^{\eta }\mathbf{g} &\mathbf{=}&e^{\psi (x^{k})}{dx^{i}\otimes dx^{i}}-\
^{0}h_{3}(x^{k})\mathbf{e}^{3}{\otimes }\mathbf{e}^{3}+h_{4}(x^{k},t)\mathbf{%
e}^{4}{\otimes }\mathbf{e}^{4},  \notag \\
\mathbf{e}^{3} &=&dt+w_{i}(x^{k},t)dx^{i},\mathbf{e}%
^{4}=dy^{4}+n_{i}(x^{k},t)dx^{i},  \label{genans2}
\end{eqnarray}%
where $g_{1}=g_{2}=e^{\psi (x^{k})},$ with $\psi (x^{k})$ being a solution
of (\ref{4ep1a}) for any given $\Upsilon _{4}(x^{k}).$ The function $%
h_{4}(x^{k},t)$ must satisfy the equation (\ref{4ep2a}) which for $%
h_{3}^{\ast }=0 $ is just
\begin{equation*}
h_{4}^{\ast \ast }-\frac{\left( h_{4}^{\ast }\right) ^{2}}{2h_{4}}-2\
^{0}h_{3}h_{4}\Upsilon _{2}(x^{k},t)=0.
\end{equation*}%
The N--connection coefficients are
\begin{equation*}
w_{i}=-\partial _{i}\widetilde{\phi }/\widetilde{\phi }^{\ast },~n_{i}=\
^{1}n_{k}\left( x^{i}\right) +\ ^{2}n_{k}\left( x^{i}\right) \int [1/(\sqrt{%
|h_{4}|})^{3}]dt,
\end{equation*}%
when $\widetilde{\phi }=\ln |h_{4}^{\ast }/\sqrt{|\ ^{0}h_{3}h_{4}|}|.$

The Levi--Civita configurations for solutions (\ref{genans2}) are selected
by the conditions (\ref{lcconstr1}) which, for this case, are satisfied if
\begin{equation*}
\ ^{2}n_{k}\left( x^{i}\right) =0\ \mbox{ and }\partial _{i}\
^{1}n_{k}=\partial _{k}\ ^{1}n_{i},
\end{equation*}%
and
\begin{eqnarray*}
\left( w_{i}[\widetilde{\phi }]\right) ^{\ast }+w_{i}[\widetilde{\phi }%
]\left( h_{4}[\widetilde{\phi }]\right) ^{\ast }+\partial _{i}h_{4}[%
\widetilde{\phi }]=0, && \\
\partial _{i}\ w_{k}[\widetilde{\phi }]=\partial _{k}\ w_{i}[\widetilde{\phi
}]. &&
\end{eqnarray*}%
Such conditions are similar to (\ref{auxc1}) but for a different relation of
v--coefficients of metric to another type of generating function $\widetilde{%
\phi }.$ They are always satisfied for cosmological solutions with $%
\widetilde{\phi }=\widetilde{\phi }(t)$ or if $\widetilde{\phi }=const$ (in
the last case $w_{i}(x^{k},t)$ can be any functions as follows from (\ref%
{4ep3a}) with zero $\beta $ and $\alpha _{i},$ see (\ref{auxphi})).

\subsubsection{Type 4: Solutions with $\protect\phi =const$}

If we fix $\phi =\phi _{0}=const$ in (\ref{auxphi}), but $h_{3}^{\ast }\neq
0 $ and $h_{4}^{\ast }\neq 0,$ we can express the general solutions of (\ref%
{4ep1a})--(\ref{4ep4a}) in the form
\begin{eqnarray}
\ ^{\eta }\mathbf{g} &\mathbf{=}&e^{\psi (x^{k})}{dx^{i}\otimes dx^{i}}-\
^{0}h^{2}\ \left[ f^{\ast }\left( x^{i},t\right) \right] ^{2}|\varsigma
_{\Upsilon }\left( x^{i},t\right) |\mathbf{e}^{3}{\otimes }\mathbf{e}^{3}
\notag \\
&& +f^{2}\left( x^{i},t\right) \mathbf{e}^{4}{\otimes }\mathbf{e}^{4},
\notag \\
\mathbf{e}^{3} &=&dt+w_{i}(x^{k},t)dx^{i},\ \mathbf{e}^{4}=dy^{4}+n_{k}%
\left( x^{i},t\right) dx^{i},  \label{genans3}
\end{eqnarray}%
where $^{0}h=const,$ $g_{1}=g_{2}=e^{\psi (x^{k})},$ with $\psi (x^{k})$
being a solution of (\ref{4ep1a}) for any given $\Upsilon _{4}(x^{k}),$ and $%
\varsigma _{\Upsilon }\left( x^{i},t\right) =\varsigma _{4[0]}\left(
x^{i}\right) -\frac{h_{0}^{2}}{16}\int \Upsilon _{2}(x^{k},t)[f^{2}\left(
x^{i},t\right) ]^{2}dt.$ The N--connection coefficients $%
N_{i}^{3}=w_{i}(x^{k},t)$ and $N_{i}^{4}=n_{i}(x^{k},t)$ are
\begin{equation}
w_{i}=-\frac{\partial _{i}\varsigma _{\Upsilon }\left( x^{k},t\right) }{%
\varsigma _{\Upsilon }^{\ast }\left( x^{k},t\right) }  \label{gensol1w}
\end{equation}%
and
\begin{equation}
n_{k}=\ ^{1}n_{k}\left( x^{i}\right) +\ ^{2}n_{k}\left( x^{i}\right) \int
\frac{\left[ f^{\ast }\left( x^{i},t\right) \right] ^{2}}{\left[ f\left(
x^{i},t\right) \right] ^{2}}\varsigma _{\Upsilon }\left( x^{i},t\right) dt.
\label{gensol1n}
\end{equation}

We must take $\varsigma _{4[0]}\left( x^{i}\right) =\pm 1$ if $\varsigma
_{\Upsilon }\left( x^{i},t\right) =\pm 1$ for $\Upsilon _{2}\rightarrow 0.$
In such a case, the functions $h_{3}=-\ ^{0}h^{2}\ \left[ f^{\ast }\left(
x^{i},t\right) \right] ^{2}$ and $h_{4}=f^{2}\left( x^{i},t\right) $ satisfy
the equation (\ref{4ep2a}) written in the form $\sqrt{|h_{3}|}=\ ^{0}h(\sqrt{%
|h_{4}|})^{\ast },$ which is compatible with the condition $\phi =\phi _{0}.$

The subclass of solutions for the Levi--Civita connection with ansatz of
type (\ref{genans3}) is subjected additionally to the conditions (\ref%
{lcconstr1}). For instance, we can chose that $\ ^{2}n_{k}\left(
x^{i}\right) =0$ and $\ ^{1}n_{k}\left( x^{i}\right) $ are any functions
satisfying the conditions $\ \partial _{i}\ ^{1}n_{k}=\partial _{k}\
^{1}n_{i}.$ The constraints on values $w_{i}=-\partial _{i}\varsigma
_{\Upsilon }/\varsigma _{\Upsilon }^{\ast }$ result in constraints on $%
\varsigma _{\Upsilon },$ which is determined by $\Upsilon _{2}$ and $f,$
\begin{eqnarray}
\left( w_{i}[\varsigma _{\Upsilon }]\right) ^{\ast }+w_{i}[\varsigma
_{\Upsilon }]\left( h_{4}[\varsigma _{\Upsilon }]\right) ^{\ast }+\partial
_{i}h_{4}[\varsigma _{\Upsilon }]=0, &&  \notag \\
\partial _{i}\ w_{k}[\varsigma _{\Upsilon }]=\partial _{k}\ w_{i}[\varsigma
_{\Upsilon }], &&  \label{auxc3}
\end{eqnarray}%
where, for instance, we denoted by $h_{4}[\varsigma _{\Upsilon }]$ the
functional dependence on $\varsigma _{\Upsilon }.$ Such conditions are
always satisfied for cosmological solutions with $f=f(t).$ For $\widehat{%
\mathbf{D}},$ if $\ \Upsilon _{2}=0$ and $\phi =const,$ the coefficients $%
w_{i}(x^{k},t)$ can be arbitrary functions (we can fix $\varsigma _{\Upsilon
}=1,$ which does not impose a functional dependence of $w_{i}$ on $\varsigma
_{\Upsilon })$ as follows from (\ref{4ep3a}) with zero $\beta $ and $\alpha
_{i},$ see (\ref{auxphi}). To generate solutions for $\nabla $ such $w_{i}$
must be additionally constrained following formulas (\ref{auxc3})
re--written for $w_{i}[\varsigma _{\Upsilon }]\rightarrow w_{i}(x^{k},t)$
and $h_{4}[\varsigma _{\Upsilon }]\rightarrow h_{4}\left( x^{i},t\right).$

We note that any solution $\mathbf{g}=\{g_{\alpha ^{\prime }\beta ^{\prime
}}(u^{\alpha ^{\prime }})\}$ of the Einstein equations (\ref{cdeinst})
and/or (\ref{einsteq}) with Killing symmetry $\partial /\partial y$ (for
local coordinates in the form $y^{3}=t$ and $y^{4}=y)$ can be parametrized
in a form derived in this section. Using frame transforms of type $e_{\alpha
}=e_{\ \alpha }^{\alpha ^{\prime }}e_{\alpha ^{\prime }},$ with $\mathbf{g}%
_{\alpha \beta }=e_{\ \alpha }^{\alpha ^{\prime }}e_{\ \beta }^{\beta
^{\prime }}g_{\alpha ^{\prime }\beta ^{\prime }},$ for any $\mathbf{g}%
_{\alpha \beta }$ (\ref{gsol}), we relate the class of such (inhomogeneous)
cosmological solutions, for instance, to the family of metrics of type (\ref%
{genans}).\footnote{%
We have to solve certain systems of quadratic algebraic equations and define
some $e_{\ \alpha }^{\alpha ^{\prime }}(u^{\beta }),$ choosing a convenient
system of coordinates $u^{\alpha ^{\prime }}=u^{\alpha ^{\prime }}(u^{\beta
}).$}

Following recent developments from Ref. \cite{vgsolhd}, we can construct
'non--Killing' solutions depending on all coordinates. Such general classes
of solutions can be parametrized in the form
\begin{eqnarray}
\mathbf{g} &\mathbf{=}&+g_{i}(x^{k}){dx^{i}\otimes dx^{i}}+\omega
^{2}(x^{j},t,y)h_{a}(x^{k},t)\mathbf{e}^{a}{\otimes }\mathbf{e}^{a},  \notag
\\
\mathbf{e}^{3} &=&dy^{3}+w_{i}(x^{k},t)dx^{i},\mathbf{e}%
^{4}=dy^{4}+n_{i}(x^{k},t)dx^{i},  \label{ansgensol}
\end{eqnarray}%
for any $\omega $ for which $\mathbf{e}_{k}\omega =\partial _{k}\omega
+w_{k}\omega ^{\ast }+n_{k}\partial \omega /\partial y=0,$ when (\ref%
{ansgensol}) with $\omega ^{2}=1$ is of type (\ref{gsol}).

Finally, we note that the metrics constructed above define general
cosmological solutions of Einstein equations for any type of sources $%
\varkappa T_{\beta \delta }$ which can be parametrized\footnote{%
using chains of nonholonomic frame transforms, this is possible for 'almost'
all physically important energy--momentum tensors} in a formally
diagonalized form $\Upsilon _{\gamma }$ (\ref{source}), with respect to a
nonholonomic frame of reference.

\section{Examples of Cosmological Models with Off--Di\-agonal Metrics}

\label{s4} The goal of this section is to analyze explicit examples of
generic off--diagonal cosmological solutions with $\Upsilon _{4}=0$ but, in
general, with non--vanishing $\Upsilon _{2}(t).$ Such solutions are
constructed as examples of metrics (\ref{gsol1}), (\ref{genans1}), (\ref%
{genans2}) and (\ref{genans3}). The new classes of cosmological metrics have
respective 'diagonal' limits to conformal and/or frame transforms of metrics
(\ref{frw1}), (\ref{bianchim}), (\ref{kasner}), (\ref{godele}).

We consider a particular parametrization of metrics of type (\ref{gsol})
when
\begin{eqnarray}
\ ^{\eta }\mathbf{g} &\mathbf{=}&\eta _{i}(x^{k},t)\ ^{\circ }g_{i}(x^{k},t){%
dx^{i}\otimes dx^{i}}+\eta _{a}(t)\ ^{\circ }h_{a}(t)\mathbf{e}^{a}{\otimes }%
\mathbf{e}^{a},  \label{anspolariz} \\
\mathbf{e}^{3} &=&dt+\eta _{i}^{3}(t)\ ^{\circ }w_{i}(t)dx^{i},\ \mathbf{e}%
^{4}=dy^{4}+\eta _{i}^{4}(t)\ ^{\circ }n_{i}(t)dx^{i},  \notag
\end{eqnarray}%
or (for an alternative parametrization)
\begin{equation*}
\mathbf{e}^{3}=dt+\left[ \eta _{i}^{3}(t)+\ ^{\circ }w_{i}(t)\right]
dx^{i},\ \mathbf{e}^{4}=dy^{4}+\left[ \eta _{i}^{4}(t)+\ ^{\circ }n_{i}(t)%
\right] dx^{i},
\end{equation*}%
when (for some constructions) $g_{1}=\eta _{1}(x^{k},t)\ ^{\circ
}g_{1}(x^{k},t)=1$ and $g_{2}=\eta _{1}(x^{k},t)\ ^{\circ }g_{2}(x^{k},t)=1$
are trivial solutions of (\ref{eq1}), and (\ref{4ep1a}), with $\Upsilon
_{4}=0.$ For homogeneous configurations, we can always introduce such
coordinates when the coefficients of metrics do not depend on space
variables (a formal presence of radial and angular variables may exist, for
instance, in spherical coordinates, like in the case of FRW metric (\ref{frw}%
)) The polarization functions $\eta _{a}(t)$ and $\eta _{i}^{a}(t)$ can be
chosen in a form which allows us to generate homogeneous configurations
(i.e. solutions not depending on space coordinates) with $\ h_{a}(t)=\eta
_{a}(t)\ ^{\circ }h_{a}(t)$ and $\ w_{i}(t)=\eta _{i}^{3}(t)\ ^{\circ
}w_{i}(t),\ n_{i}(t)=\eta _{i}^{4}(t)\ ^{\circ }n_{i}(t).$

\subsection{Nonholonomic FRW, Bianchi, Kasner and G\"{o}del type
configurations}

\label{ofanissol}

\subsubsection{N--anholonomic FRW generalizations}

\paragraph{Solutions of type 1:}

We chose $^{\circ }g_{i}=1,\ ^{\circ }h_{3}(t)=-a^{-2}(t),\ ^{\circ
}h_{4}=1, $ where $\ a(t)$ is determined by equations (\ref{fr1}) and (\ref%
{fr2}) for the FRW model. A class of anistoropic and inhomogeneous solutions
parametrized by metrics of type (\ref{gsol1}) is generated by data%
\begin{eqnarray*}
g_{1} &=&g_{2}=1,\ \ h_{3}=\pm \ \frac{|\phi ^{\ast }(x^{i},t)|}{\Upsilon
_{2}(x^{i},t)},\  \\
\ h_{4} &=&\ ^{0}h_{4}(x^{k})\pm \ 2\int \frac{(\exp [2\ \phi
(x^{k},t)])^{\ast }}{\Upsilon _{2}(x^{i},t)}dt,\  \\
w_{i} &=&-\partial _{i}\phi /\phi ^{\ast },\ n_{i}=\ ^{1}n_{k}\left(
x^{i}\right) +\ ^{2}n_{k}\left( x^{i}\right) \int [h_{3}/(\sqrt{|h_{4}|}%
)^{3}]dt.
\end{eqnarray*}%
Off--diagonal cosmological metrics depending only on variable $t$ can be
extracted by gravitational polarizations
\begin{eqnarray}
\eta _{i} &=&1,\eta _{3}=a^{2}(t)|\phi ^{\ast }(t)|/\Upsilon _{2}(t),\eta
_{4}=\ 1\pm \ 2\int \frac{(\exp [2\ \phi (t)])^{\ast }}{\Upsilon _{2}(t)}dt,
\notag \\
\eta _{i}^{3}(t) &=&0,\eta _{i}^{4}(t)=\ ^{1}n_{k}+\ ^{2}n_{k}\int [\eta
_{3}(t)/(\sqrt{|\eta _{4}(t)|})^{3}]dt,  \label{aux1}
\end{eqnarray}%
with some $\ ^{1}n_{k}=const,\ ^{2}n_{k}=const,$ when $w_{i}(t)=\eta
_{i}^{3}(t)+\ ^{\circ }w_{i}(t)=0,$ for $\ ^{\circ }w_{i}(t)=0,$ and $%
n_{i}(t)=\eta _{i}^{4}(t)+\ ^{\circ }n_{i}(t),$ for $\ ^{\circ }n_{i}(t)=0.$
The factor $a^{2}(t)$ can be included into a generating function $\phi (t),$
or into a source $\Upsilon _{2}(t)$ (we can say that it is nonholonomically
modelled by such a generating function, alternatively, effective source).
The off--diagonal metric
\begin{eqnarray}
\ ^{\eta }\mathbf{g} &\mathbf{=}&{dx^{1}\otimes dx^{1}+dx^{2}\otimes dx^{2}}+%
\frac{|\phi ^{\ast }(t)|}{\Upsilon _{2}(t)}dt{\otimes }dt+  \label{sol11} \\
&&\left[ 1\pm \ 2\int \frac{(\exp [2\ \phi (t)])^{\ast }}{\Upsilon _{2}(t)}dt%
\right] \left[ dy^{4}+\eta _{i}^{4}(t)dx^{i}\right] {\otimes }\left[
dy^{4}+\eta _{i}^{4}(t)dx^{i}\right] ,  \notag
\end{eqnarray}%
with $\eta _{i}^{4}(t)$ computed following formulas (\ref{aux1}) defines a
class of cosmological solutions of the Einstein equations (\ref{cdeinst})
for $\widehat{\mathbf{D}}$ with a source $\ \Upsilon _{\ \ \beta }^{\alpha
}=diag[\Upsilon _{\gamma };\Upsilon _{1}=\Upsilon _{2}=\Upsilon
_{2}(t);\Upsilon _{3}=\Upsilon _{4}=0].$ We can extract Levi--Civita
configurations if we fix, for instance, \ the integration constants $\
^{1}n_{k}=\ ^{2}n_{k}=0.$

For gravitational polarizations with certain smooth limits
\begin{equation*}
\frac{|\phi ^{\ast }(t)|}{\Upsilon _{2}(t)}\rightarrow -a^{-2}(t),\int \frac{%
(\exp [2\ \phi (t)])^{\ast }}{\Upsilon _{2}(t)}dt\rightarrow 0,\eta
_{i}^{4}(t)\rightarrow 0
\end{equation*}%
(this can be satisfied by a corresponding choosing of functions $\phi
(t),\Upsilon _{2}(t)$ and integration constants $\ ^{1}n_{k},\ ^{2}n_{k}),$
the solutions (\ref{sol11}) transform not just into the FRW metric (\ref{frw}%
) but into its conformal transform (with multiplication on factor $a^{-2}(t)$%
). Such off--diagonal cosmological solutions have certain similarities (for
small nonholonomic deformations) to conformally transformed FRW solutions if
$\phi (t),\Upsilon _{2}(t)$ are fixed to mimic $\ a(t)$ as in equations (\ref%
{fr1}) and (\ref{fr2}) for the Hubble parameter.

\vskip4pt

\paragraph{Solutions of type 2:}

As for the type 1, we also take $^{\circ }g_{i}=1,\ ^{\circ
}h_{3}(t)=-a^{-2}(t),\ ^{\circ }h_{4}=1$ but generate anistoropic and
inhomogeneous solutions parametrized by metrics of type (\ref{genans1})
\begin{eqnarray}
\ ^{\eta }\mathbf{g} &\mathbf{=}&{dx^{1}\otimes dx^{1}+dx^{2}\otimes dx^{2}}%
+h_{3}(x^{k},t)\mathbf{e}^{3}{\otimes }\mathbf{e}^{3}+\ ^{0}h_{4}(x^{k})%
\mathbf{e}^{4}{\otimes }\mathbf{e}^{4},  \label{sol12} \\
\mathbf{e}^{3} &=&dt+w_{i}(x^{k},t)dx^{i},\ \mathbf{e}^{4}=dy^{4}+\left[ \
^{1}n_{k}\left( x^{i}\right) +\ ^{2}n_{k}\left( x^{i}\right) \int h_{3}dt%
\right] dx^{i},  \notag
\end{eqnarray}%
for arbitrary generating functions $h_{3}(x^{k},t),w_{i}(x^{k},t),\
^{0}h_{4}(x^{k})$ and integration functions $\ ^{1}n_{k}\left( x^{i}\right) $
and $\ ^{2}n_{k}\left( x^{i}\right) ,$ when $\Upsilon _{2}=0.$ The
polarization functions are
\begin{eqnarray*}
\eta _{i} &=&1,\eta _{3}=a^{2}(t)h_{3}(x^{k},t),\eta _{4}=\ \
^{0}h_{4}(x^{k}), \\
\eta _{i}^{3} &=&w_{i}(x^{k},t),\eta _{i}^{4}=\ ^{1}n_{k}\left( x^{i}\right)
+\ ^{2}n_{k}\left( x^{i}\right) \int h_{3}dt,
\end{eqnarray*}%
with $\ ^{\circ }w_{i}(t)=0$ and $\ ^{\circ }n_{i}(t)=0.$ Such cosmological
solutions are anisotropic and inhomogeneous and constructed as nonholonomic
deformations of a conformally transformed (with multiplication on factor $%
a^{-2}(t)$) FRW metric (\ref{frw}). We can prescribe polarization functions $%
\eta _{3}(x^{k},t)$ when $h_{3}=\eta _{3}\ ^{\circ }h_{3}(t)\rightarrow $ $%
-a^{-2}(t)$ for $\eta _{3}\rightarrow 1.$

The conditions (\ref{lcconstr1}) selecting Levi--Civita configurations
transform into equations
\begin{eqnarray}
\ ^{2}n_{k}\left( x^{i}\right) &=&0\ \mbox{ and }\partial _{i}\
^{1}n_{k}=\partial _{k}\ ^{1}n_{i},  \label{lccond12} \\
w_{i}^{\ast }+\partial _{i}\ ^{0}h_{4} &=&0\mbox{ and }\partial _{i}\
w_{k}=\partial _{k}\ w_{i}.  \notag
\end{eqnarray}%
Such constraints can be satisfied for any generating function $%
h_{3}(x^{k},t) $ but impose additional constraints on N--coefficients $%
w_{i}(x^{k},t).$

Off--diagonal metrics of type (\ref{sol12}) can be generated to be only with
time like dependence of coefficients, \ when $h_{3}=h_{3}(t),w_{i}=w_{i}(t)$
and $n_{i}(t)$ are determined with some constant values of $\ ^{0}h_{4},\ \
^{1}n_{k},\ ^{2}n_{k}.$ Such conditions are for the Levi--Civita
configurations if $w_{i}=const.$ This defines vacuum solutions of the
Einstein equations. They transform conformally and nonholonomically a FRW
universe into certain vacuum Einstein \ configurations which (in this
particular case) can be diagonalized by coordinate transforms.

\vskip4pt

\paragraph{Solutions of type 3:}

It is not possible to construct off--diagonal generalizations of type (\ref%
{genans2}), with coordinate $y^{3}=t,$ for FRW universes, when $h_{3}=-\
^{0}h_{3}(x^{k})$ does not depend on $t$ and $^{\circ }g_{i}=1,\ ^{\circ
}h_{3}(t)=-a^{-2}(t),\ ^{\circ }h_{4}=1.$ For such solutions, we can not
obtain $h_{3}\sim a^{-2}(t)$ $\ $in the limit of trivial polarizations.
Considering inhomogeneous metrics of type (\ref{ansgensol}) with $\omega
=\omega (x^{i},t,y),$ where $a^{-2}(t)$ can be included into $\omega ^{2},$
we model general inhomogeneous solutions obtained via ''vertical'' conformal
transforms and further nonholonomic deformations. The metrics are of type
\begin{eqnarray}
\ ^{\eta }\mathbf{g} &\mathbf{=}&{dx^{1}\otimes dx^{1}+dx^{2}\otimes dx^{2}+}
\label{sol13} \\
&&a^{-2}(t)\widetilde{\omega }^{2}(x^{i},t,y)\left[ -\ ^{0}h_{3}(x^{k})%
\mathbf{e}^{3}{\otimes }\mathbf{e}^{3}+h_{4}(x^{k},t)\mathbf{e}^{4}{\otimes }%
\mathbf{e}^{4}\right] ,  \notag \\
\mathbf{e}^{3} &=&dt+w_{i}(x^{k},t)dx^{i},\mathbf{e}%
^{4}=dy^{4}+n_{i}(x^{k},t)dx^{i},  \notag
\end{eqnarray}%
where $\omega =\widetilde{\omega }/a$ is a solution of $\partial _{k}\omega
+w_{k}\omega ^{\ast }+n_{k}\partial \omega /\partial y=0.$ The equation (\ref%
{4ep2a}) for $h_{3}^{\ast }=0$ transforms into
\begin{equation}
h_{4}^{\ast \ast }-\frac{\left( h_{4}^{\ast }\right) ^{2}}{2h_{4}}-2\
^{0}h_{3}h_{4}\Upsilon _{2}(x^{k},t)=0.  \label{veq}
\end{equation}%
Integrating two times on $t,$ we can define $h_{4}(x^{k},t)$ for any given $%
\ ^{0}h_{3}(x^{k})$ and $\Upsilon _{2}(x^{k},t).$ Then we can compute the
N--connection coefficients following formulas
\begin{equation}
w_{i}=-\partial _{i}\widetilde{\phi }/\widetilde{\phi }^{\ast },~n_{i}=\
^{1}n_{k}\left( x^{i}\right) +\ ^{2}n_{k}\left( x^{i}\right) \int [1/(\sqrt{%
|h_{4}|})^{3}]dt,  \label{ncoef}
\end{equation}%
for $\widetilde{\phi }=\ln |h_{4}^{\ast }/\sqrt{|\ ^{0}h_{3}h_{4}|}|.$

The Levi--Civita configurations are extracted from the set of solutions (\ref%
{sol13}) if
\begin{equation*}
\ ^{2}n_{k}\left( x^{i}\right) =0\ \mbox{ and }\partial _{i}\
^{1}n_{k}=\partial _{k}\ ^{1}n_{i},
\end{equation*}%
and
\begin{eqnarray*}
\left( w_{i}[\widetilde{\phi }]\right) ^{\ast }+w_{i}[\widetilde{\phi }%
]\left( h_{4}[\widetilde{\phi }]\right) ^{\ast }+\partial _{i}h_{4}[%
\widetilde{\phi }]=0, && \\
\partial _{i}\ w_{k}[\widetilde{\phi }]=\partial _{k}\ w_{i}[\widetilde{\phi
}]. &&
\end{eqnarray*}%
Such conditions are always satisfied for cosmological solutions with $%
\widetilde{\phi }=\widetilde{\phi }(t)$ or if $\widetilde{\phi }=const$ (in
the last case $w_{i}(x^{k},t)$ can be any functions as follows from (\ref%
{4ep3a}) with zero $\beta $ and $\alpha _{i},$ see (\ref{auxphi}); in such
cases, we must take $\Upsilon _{2}=0).$

The first subclass of metrics (\ref{sol13}) (with solutions depending only
on $t,$ with nontrivial $\omega (t,y),$ for constant $\ ^{0}h_{3}$ and$\
^{1}n_{k};$ $\ ^{2}n_{k}=0$ and nontrivial $\Upsilon _{2}(t))$ are
parametrized in the form
\begin{eqnarray}
\ ^{\eta }\mathbf{g} &\mathbf{=}&{dx^{1}\otimes dx^{1}+dx^{2}\otimes dx^{2}+}%
a^{-2}(t)\widetilde{\omega }^{2}(t,y)\times  \label{sol13a} \\
&&\left[ -\ ^{0}h_{3}dt{\otimes }dt+h_{4}(t)(dy^{4}+\ ^{1}n_{i}dx^{i}){%
\otimes }(dy^{4}+\ ^{1}n_{i}dx^{i})\right] ,  \notag
\end{eqnarray}%
where $h_{4}(t)$ is any solution of $h_{4}^{\ast \ast }-\frac{\left(
h_{4}^{\ast }\right) ^{2}}{2h_{4}}-2\ ^{0}h_{3}h_{4}\Upsilon _{2}(t)=0.$

The second subclass of metrics (\ref{sol13}) contains arbitrary functions $%
w_{i}(t)$ but for $\Upsilon _{2}(t)=0$ and $h_{4}(t)$ is any solution of $%
h_{4}^{\ast \ast }-\left( h_{4}^{\ast }\right) ^{2}/(2h_{4})=0.$ For
constant $\ ^{0}h_{3}$ and$\ ^{1}n_{k}$ and $\ ^{2}n_{k}=0,$ we get
\begin{eqnarray}
\ ^{\eta }\mathbf{g} &\mathbf{=}&{dx^{1}\otimes dx^{1}+dx^{2}\otimes dx^{2}+}%
a^{-2}(t)\widetilde{\omega }^{2}(t,y)\times  \label{sol13b} \\
&&\lbrack -\ ^{0}h_{3}(dt+w_{i}(t)dx^{i}){\otimes }(dt+w_{i}(t)dx^{i})
\notag \\
&& +h_{4}(t)(dy^{4}+\ ^{1}n_{i}dx^{i}){\otimes }(dy^{4}+\ ^{1}n_{i}dx^{i})].
\notag
\end{eqnarray}

Both types of solutions (\ref{sol13a}) and (\ref{sol13b}) are for the
Levi--Civita connection. They can be generalized for the canonical
d--connection by introducing nontrivial constants $\ ^{2}n_{k}$ in the
formulas for N--connection coefficients, $N_{i}^{4}=n_{i}(t),$ see (\ref%
{ncoef}), with $h_{4}(t)$ respectively computed as in equation (\ref{veq}),
when $\Upsilon _{2}=\Upsilon _{2}(t)$ and or $\Upsilon _{2}=0.$

For the metric (\ref{sol13}), the polarization functions are$\
^{0}h_{3}(x^{k})$
\begin{eqnarray*}
\eta _{i} &=&1,\eta _{3}=\widetilde{\omega }^{2}(x^{i},t,y)\
^{0}h_{3}(x^{k}),\eta _{4}=a^{-2}(t)\widetilde{\omega }^{2}(x^{i},t,y)\ \
h_{4}(x^{k},t), \\
\eta _{i}^{3} &=&w_{i}(x^{k},t),\eta _{i}^{4}=\ ^{1}n_{k}\left( x^{i}\right)
+\ ^{2}n_{k}\left( x^{i}\right) \int h_{3}dt,
\end{eqnarray*}%
with $\ ^{\circ }w_{i}(t)=0$ and $\ ^{\circ }n_{i}(t)=0.$ We have to
eliminate respectively the dependencies on $x^{i}$ in these formulas in
order to compute the gravitational polarizations for the metrics (\ref%
{sol13a}) and (\ref{sol13b}). For simplicity, we omit the formulas for this
class of solutions. \ Such classes of metrics limit to the conformally
transformed FRW one (with conformal factor $a^{-2}(t))$ if $\eta _{\alpha
}\rightarrow 1$ and $\eta _{i}^{a}\rightarrow 0.$

\vskip4pt

\paragraph{Solutions of type 4:}

If we fix $\phi =\phi _{0}=const$ in (\ref{auxphi}), but $h_{3}^{\ast }\neq
0 $ and $h_{4}^{\ast }\neq 0,$ we have to reparametrize the sets of
generating and integration functions in a different form. The solutions (\ref%
{genans3}) with $^{\circ }g_{i}=1,\ ^{\circ }h_{3}(t)=-a^{-2}(t),\ ^{\circ
}h_{4}=1$ are of type
\begin{eqnarray}
\ ^{\eta }\mathbf{g} &=&{dx^{1}\otimes dx^{1}+dx^{2}\otimes dx^{2}}  \notag
\\
&& -\ ^{0}h^{2}\ \left[ f^{\ast }\left( x^{i},t\right) \right]
^{2}|\varsigma _{\Upsilon }\left( x^{i},t\right) |\mathbf{e}^{3}{\otimes }%
\mathbf{e}^{3}+f^{2}\left( x^{i},t\right) \mathbf{e}^{4}{\otimes }\mathbf{e}%
^{4},  \notag \\
\mathbf{e}^{3} &=&dt+w_{i}(x^{k},t)dx^{i},\ \mathbf{e}^{4}=dy^{4}+n_{k}%
\left( x^{i},t\right) dx^{i},  \label{sol14}
\end{eqnarray}%
where
\begin{eqnarray*}
\varsigma _{\Upsilon } &=&1-\frac{h_{0}^{2}}{16}\int \Upsilon
_{2}(x^{k},t)[f^{2}\left( x^{i},t\right) ]^{2}dt, \\
w_{i} &=&-\partial _{i}\varsigma _{\Upsilon }\left( x^{k},t\right)
/\varsigma _{\Upsilon }^{\ast }\left( x^{k},t\right) , \\
n_{k} &=&\ ^{1}n_{k}\left( x^{i}\right) +\ ^{2}n_{k}\left( x^{i}\right) \int
\frac{\left[ f^{\ast }\left( x^{i},t\right) \right] ^{2}}{\left[ f\left(
x^{i},t\right) \right] ^{2}}\varsigma _{\Upsilon }\left( x^{i},t\right) dt.
\end{eqnarray*}

The Levi--Civita configurations are chosen when $\ ^{2}n_{k}=0$ and $\
^{1}n_{k}\left( x^{i}\right) $ are any functions with $\ \partial _{i}\
^{1}n_{k}=\partial _{k}\ ^{1}n_{i}.$ The constraints for $w_{i}=-\partial
_{i}\varsigma _{\Upsilon }/\varsigma _{\Upsilon }^{\ast }$ are similar to (%
\ref{auxc3}).

We obtain from (\ref{sol14}) generic off--diagonal homogeneous cosmological
solutions if we consider metrics with dependence only on $t.$ They can be
written in the form
\begin{eqnarray}
\ ^{\eta }\mathbf{g} &\mathbf{=}&{dx^{1}\otimes dx^{1}+dx^{2}\otimes dx^{2}}%
-\ ^{0}h^{2}\ \left[ f^{\ast }\left( t\right) \right] ^{2}|\varsigma
_{\Upsilon }\left( t\right) |\mathbf{e}^{3}{\otimes }\mathbf{e}%
^{3}+f^{2}\left( t\right) \mathbf{e}^{4}{\otimes }\mathbf{e}^{4},  \notag \\
\mathbf{e}^{3} &=&dt+w_{i}(t)dx^{i},\ \mathbf{e}^{4}=dy^{4}+n_{k}\left(
t\right) dx^{i},  \label{sol14a}
\end{eqnarray}%
where, for constant $\ ^{1}n_{k},\ ^{2}n_{k}$ and $\ ^{0}h,$ and $\varsigma
_{\Upsilon }\left( t\right) =1-\frac{h_{0}^{2}}{16}\int \Upsilon
_{2}(t)[f^{2}\left( t\right) ]^{2}dt,$
\begin{equation*}
w_{i}=-\partial _{i}\varsigma _{\Upsilon }\left( t\right) /\varsigma
_{\Upsilon }^{\ast }\left( t\right) ,\ n_{k}=\ ^{1}n_{k}+\ ^{2}n_{k}\int
\frac{\left[ f^{\ast }\left( x^{i},t\right) \right] ^{2}}{\left[ f\left(
x^{i},t\right) \right] ^{2}}\varsigma _{\Upsilon }\left( t\right) dt.
\end{equation*}%
For simplicity, we omit the explicit formulas for gravitational
polarizations (they can be computed similarly to those given for above
examples).

Taking $\ ^{2}n_{k}=0$ for (\ref{sol14a}) and imposing constraints of type (%
\ref{auxc3}) for $w_{i}(t),$ we generate new classes of cosmological
solutions in general relativity. For trivial gravitational polarizations and
vanishing N--connection coefficients, such metrics result in the conformal
transform (with factor $a^{-2}(t)$) of the FRW metric (\ref{frw}).

\subsubsection{Locally anisotropic Bianchi spacetimes}

We speculate how a Bianchi metric $\ _{B}\mathbf{g}_{\alpha ^{\prime \prime
}\beta ^{\prime \prime }}$ (\ref{bianchim}) can be generalized into locally
anisotropic solutions. Taking any set of coefficients $_{B}g_{\underline{%
\alpha }\underline{\beta }}(t),$ we have to construct certain frame
transform, $\ _{B}^{\circ }\mathbf{g}_{\alpha \beta }=\ \ _{B}e_{\ \alpha
^{\prime \prime }}^{\underline{\alpha }}\ \ _{B}e_{\ \beta ^{\prime \prime
}}^{\underline{\beta }}\ \ _{B}g_{\underline{\alpha }\underline{\beta }},$%
\footnote{%
we have to solve certain quadratic algebraic equations in order to define
frame coefficients depending on coordinate $t,$ or being some constants}
when $_{B}^{\circ }\mathbf{g}_{\alpha \beta }$ is parametrized as a prime
metric
\begin{equation}
\ _{B}^{\circ }\mathbf{g}\mathbf{=}\ ^{\circ }g_{i}{dx^{i}\otimes dx^{i}}+\
^{\circ }h_{a}(t)\mathbf{e}^{a}{\otimes }\mathbf{e}^{a},\ \mathbf{e}%
^{3}=dt+\ ^{\circ }w_{i}(t)dx^{i},\ \mathbf{e}^{4}=dy^{4}+\ ^{\circ
}n_{i}(t)dx^{i}.  \notag
\end{equation}%
Introducing corresponding gravitational $\eta $--polarizations, we construct
nonholonomic deformations $\ _{B}^{\circ }\mathbf{g\rightarrow }\ _{B}^{\eta
}\mathbf{g,}$ where the target metric can be any one of type (\ref{ansgensol}%
), or (for $\omega ^{2}=1)$ of type (\ref{gsol}). In general, such
configurations are locally anisotropic and inhomogeneous when the solutions
depend on all coordinates.

We can derive generic off--diagonal solutions with the coefficients
depending only on $t,$ or being some constants, as we proved in details in
previous sections. The main difference is that for trivial gravitational
polarizations such metrics describe frame/conformal transforms of Bianchi
spacetimes and not of conformal transforms of FRW metrics. Explicit
constructions depend on ''structure constants'' (\ref{bstrc}) for the prime
metric and are, in general, different for four types of off--diagonal
generalizations. The length of this paper does not allow us to provide such
details and analyze possible applications in cosmology.

\subsubsection{Off--diagonal Kasner type metrics}

The data for the primary metric are taken in the form $\ ^{\circ }g_{1}=1,\
^{\circ }g_{2}=t^{2(p_{3}-p_{1})},\ ^{\circ }h_{3}=-t^{-2p_{1}},\ ^{\circ
}h_{4}=t^{2(p_{2}-p_{1})}$ and $\ ^{\circ }N_{i}^{a}=0$ with constants $%
p_{1},p_{2}$ and $p_{3}$ as in the metric (\ref{kasner}). For simplicity, we
analyze in this section solutions with $p_{3}=p_{1}$ and consider an example
when a Kasner universe is generalized to locally anisotropic configurations
of type 4 characterized by gravitational polarizations%
\begin{eqnarray*}
\ \eta _{i} &=&1,\eta _{3}=\ ^{0}h^{2}\ \left[ f^{\ast }\left(
x^{i},t\right) \right] ^{2},\eta _{4}=f\left( x^{i},t\right) , \\
\eta _{i}^{3} &=&w_{i}(x^{i},t),\eta _{i}^{4}=n_{i}(x^{k},t).
\end{eqnarray*}%
For $h_{a}=\eta _{a}\ ^{\circ }h_{a}$ and $N_{i}^{a}=\eta _{i}^{a}+\ ^{\circ
}N_{i}^{a},$ the target metric is of type (\ref{genans3}) generated for $%
\Upsilon _{2}=0,$
\begin{eqnarray}
\ ^{\eta }\mathbf{g} &\mathbf{=}&{dx^{1}\otimes dx^{1}+dx^{2}\otimes dx^{2}}
\label{sol31} \\
&&-\ ^{0}h^{2}\ \left[ f^{\ast }\left( x^{i},t\right) \right] ^{2}t^{-2p_{1}}%
\mathbf{e}^{3}{\otimes }\mathbf{e}^{3}+f^{2}\left( x^{i},t\right) t^{-2p_{1}}%
\mathbf{e}^{4}{\otimes }\mathbf{e}^{4},  \notag \\
\mathbf{e}^{3} &=&dt+w_{i}(x^{k},t)dx^{i},\ \mathbf{e}^{4}=dy^{4}+n_{k}%
\left( x^{i},t\right) dx^{i},  \notag
\end{eqnarray}%
where we can consider arbitrary $w_{i}=w_{i}(x^{i},t)$ and
\begin{equation*}
\ n_{k}=\ ^{1}n_{k}(x^{i})+\ ^{2}n_{k}(x^{i})\int \frac{\left[ f^{\ast
}\left( x^{i},t\right) \right] ^{2}}{\left[ f\left( x^{i},t\right) \right]
^{2}}dt.
\end{equation*}%
The coefficients $h_{a}$ are solutions of the Einstein equations for the
canonical d--connection, see (\ref{eq2}) (and/or, equivalently, the equation
(\ref{4ep2a})) written in the form $\sqrt{|h_{3}|}=\ ^{0}h(\sqrt{|h_{4}|}%
)^{\ast }$ for arbitrary generating function $f\left( x^{i},t\right) $ if $%
p_{2}=p_{1}.$ We have to impose additional constraints on $f\left(
x^{i},t\right) $ if the last condition is not satisfied.

In the limit of trivial polarizations, the metric (\ref{sol31}) results into
a conformal transform (with factor $t^{2p_{1}})$ of Kasner solution (\ref%
{kasner}). In general, such a prime metric is not a solution of the Einstein
equations for the Levi--Civita connection but it is possible to chose
gravitational polarizations generating vacuum off--diagonal Einstein fields
even the conditions of type (\ref{kasner1}) are not satisfied. Such target
metrics may be stable (it is necessary an additional analysis of stability
properties for any explicit case).

We can eliminate dependence on space coordinates and generate solutions of
type
\begin{eqnarray*}
\ ^{\eta }\mathbf{g} &=&{dx^{1}\otimes dx^{1}+dx^{2}\otimes dx^{2}} \\
&&-\ ^{0}h^{2}\ \left[ f^{\ast }\left( t\right) \right] ^{2}t^{-2p_{1}}%
\mathbf{e}^{3}{\otimes }\mathbf{e}^{3}+f^{2}\left( t\right) t^{-2p_{1}}%
\mathbf{e}^{4}{\otimes }\mathbf{e}^{4}, \\
\mathbf{e}^{3} &=&dt+w_{i}(t)dx^{i},\ \mathbf{e}^{4}=dy^{4}+n_{k}\left(
t\right) dx^{k},
\end{eqnarray*}%
for arbitrary $w_{i}=w_{i}(t)$ and constant $\ ^{1}n_{k}$ and $\ ^{2}n_{k},$
when $n_{k}=\ ^{1}n_{k}+\ ^{2}n_{k}\int dt\left[ f^{\ast }\left( t\right) %
\right] ^{2}/\left[ f\left( t\right) \right] ^{2}.$ To extract Levi--Civita
configurations we must fix $\ ^{2}n_{k}=0$ and impose constraints of type (%
\ref{auxc3}) on $w_{i}(t)$.

In a similar form, we can construct nonholonomic deformations of the Kasner
universes of types 1-3 and/or to generalize them to solutions of type (\ref%
{ansgensol}). The corresponding target metrics may be with ''gravitational
chaos'' or constrained nonholonomically to became stable. We omit such
details in this work.

\subsubsection{Rotating cosmologies with local anisotropy}

We can chose the prime's metric data to be given by the G\"{o}del solution (%
\ref{godele}) when $\ ^{\circ }g_{i}=\ _{G}g_{i}(x),\ ^{\circ }h_{a}=\ \
_{G}h_{a}$ and $\ \ ^{\circ }N_{i}^{a}=\ _{G}N_{i}^{a},$ with local
coordinates $x^{i}=(x,z)$ and $y^{a}=(t,y).$\ Considering polarizations
\begin{eqnarray*}
\eta _{1}&=&e^{\psi x,z)},\eta _{2}=2e^{\psi (x,z)-2x},\eta _{3}=\eta
_{3}(x,z,t),\eta _{4}=\ \eta _{4}(x,z,t), \\
\eta _{i}^{3}&=&w_{i}(x,z,t),\eta _{i}^{4}=n_{k}\left( x,z,t\right) ,
\end{eqnarray*}%
for $\ g_{i}=\eta _{i}\ ^{\circ }g_{i},\ ^{\circ }h_{a}=\eta _{a}\ ^{\circ
}h_{a},\ N_{i}^{a}=\eta _{i}^{3}+\ ^{\circ }N_{i}^{a},$ when $%
N_{i}^{3}=w_{i}(x,z,t)$ and $N_{i}^{4}=n_{i}(x,z,t)$, we generate metrics of
type
\begin{eqnarray}
\ _{G}^\eta \mathbf{g} &=&\ _{G}a^{2}[e^{\psi x,z)}\left( dx{\otimes }dx+dz{%
\otimes }dz\right)  \notag \\
&&-\eta _{3}(x,z,t)(dt+w_{i}(x,z,t)dx^{i}){\otimes }(dt+w_{i}(x,z,t)dx^{i})
\notag \\
&&+\eta _{4}(x,z,t)(dy+n_{i}(x,z,t)dx^{i}){\otimes }(dy+n_{i}(x,z,t)dx^{i})].
\label{sol41}
\end{eqnarray}%
Choosing a source determined by cosmological constant $\Upsilon
_{2}=\Upsilon _{4}=\ _{G}\lambda ,$ we construct a class of type 1 exact
solutions if
\begin{eqnarray*}
\ddot{\psi}+\psi ^{\prime \prime } &=&2\ _{G}\lambda ,\ \eta _{4}^{\ast }=2\
_{G}a^{2}\ \eta _{3}\eta _{4}\ \ _{G}\lambda /\phi ^{\ast }, \\
\beta w_{i}+\alpha _{i} &=&0,\ n_{i}^{\ast \ast }+\gamma n_{i}^{\ast }=0,
\end{eqnarray*}%
with the coefficients
\begin{eqnarray*}
~\ \alpha _{i} &=&\ _{G}a^{2}\ \eta _{4}^{\ast }\partial _{i}\phi ,\ \beta
=\ _{G}a^{2}\ \eta _{4}^{\ast }\ \phi ^{\ast },\mbox{ for\  }\phi
(x,z,t)=\ln |\frac{\eta _{4}^{\ast }}{\sqrt{|\ _{G}a^{2}\ \eta _{3}\eta _{4}|%
}}|,\  \\
\gamma &=&\left( \ln |\ _{G}a^{2}\ \eta _{4}|^{3/2}/|\ _{G}a^{2}\ \eta
_{3}|\right) ^{\ast }.
\end{eqnarray*}%
Such solutions are derived directly from prime G\"{o}del metrics and have
limits to rotating universes for trivial polarizations.

In explicit form, we can model locally anisotropic and inhomogeneous models
with rotation when $h_{4}^{\ast }\neq 0;$ for $\ _{G}\lambda \neq 0,$ we get
$\phi ^{\ast }\neq 0.$ We obtain (computing consequently for a prescribed
generating function $\phi (x,z,t)$)
\begin{eqnarray}
\eta _{1} &=&e^{\psi x,z)},\eta _{2}=2e^{\psi (x,z)-2x},\eta _{3}=\pm \ \
_{G}\lambda ^{-1}\times |\phi ^{\ast }(x,z,t)|,\   \notag \\
\eta _{4} &=&\ ^{0}\eta _{4}(x,z)\pm \ 2\ _{G}\lambda ^{-1}\times \exp [2\
\phi (x,z,t)]dt,\   \label{sol41a} \\
w_{i} &=&-\partial _{i}\phi /\phi ^{\ast },  \notag \\
n_{i} &=&\ ^{1}n_{k}\left( x,z\right) +\ ^{2}n_{k}\left( x,z\right) \int
[\eta _{3}(x,z,t)/(\sqrt{|\eta _{4}(x,z,t)|})^{3}]dt.  \notag
\end{eqnarray}

The metric (\ref{sol41}) $\ $with coefficients (\ref{sol41a}) is an explicit
example of solutions of type (\ref{gsol1}) when the source is determined by
a cosmological constant. It defines a rotating cosmology additionally
imbedded into nontrivial gravitational backgrounds, which (in general) are
locally anisotropic and inhomogeneous. We can impose restrictions of type (%
\ref{lcconstr1}) and select Levi--Civita configurations. It is possible also
to construct models with anisotropic rotation when such solutions do not
depend on $x^{i},$ but only on $t,$ or with generalizations of the G\"{o}del
model determined by off--diagonal solutions of type 2-4.

\subsection{Modeling anisotropic de Sitter configurations}

We can generate cosmological solutions when the coefficients $g_{i}(x^{k},t)$
in (\ref{gsol}) depend explicitly on variable $t.$ Let us consider a
conformal factor $q(x^{k},t)$ when
\begin{eqnarray}
\ ^{q}\mathbf{g} &=&q^{2}(x^{k},t)\mathbf{[}\eta _{i}(x^{k})\ ^{\circ }g_{i}{%
dx^{i}\otimes dx^{i}}+  \notag \\
&&\eta _{3}(x^{k},t)\ ^{\circ }h_{3}(t)\mathbf{e}^{3}{\otimes }\mathbf{e}%
^{3}+\ ^{\circ }h_{4}(x^{k})\mathbf{e}^{4}{\otimes }\mathbf{e}^{4}],  \notag
\\
\mathbf{e}^{3} &=&dt+w_{i}(x^{k},t)dx^{i},\mathbf{e}%
^{4}=dy^{4}+n_{i}(x^{k},t)dx^{i}.  \label{gsolq}
\end{eqnarray}%
By straightforward computations, we can prove that the Riemann and Ricci
tensor for an arbitrary metric compatible d--connection, see details in \cite%
{vsgg}, do not change under transform $\ ^{\eta }\mathbf{g}\mathbf{=[}%
g_{ij},h_{ab},N_{i}^{a}\mathbf{]\rightarrow }\ ^{q}\mathbf{g}\mathbf{=[}%
q^{2}g_{ij},q^{2}h_{ab},N_{i}^{a}]$ if
\begin{equation}
\mathbf{e}_{i}q=\partial q/\partial x^{i}-w_{i}q^{\ast }=0.  \label{confeq}
\end{equation}%
For an ansatz of form (\ref{gsolq}), the Einstein equations for the
canonical d--connection (\ref{cdeinst}) with source (\ref{source}) is
equivalent to equations (\ref{eq1})--(\ref{eq4}) and additional equations (%
\ref{confeq}) \ for $q(x^{k},t).$

We search a subclass of inhomogeneous solutions when $q=\widetilde{q}%
(x^{k},t)a^{2}(t),$ with a prime metric $\ ^{\circ }g_{i}=1,\ ^{\circ
}h_{3}=a^{-2}(t),\ ^{\circ }h_{4}=\ ^{\circ }h_{4}(x^{k}),\ ^{\circ
}N_{i}^{a}=0,$ and $\eta $--polarizations chosen such a way that data
\begin{eqnarray*}
g_{i} &=&\eta _{i}(x^{k}),\ h_{3}=\eta _{3}(x^{k},t)\ ^{\circ }h_{3}(t),\
h_{4}=\ ^{\circ }h_{4}(x^{k}), \\
w_{i} &=&\eta _{i}^{3}(x^{k},t)+\ ^{\circ }N_{i}^{3},\ n_{i}=\eta
_{i}^{4}(x^{k},t)+\ ^{\circ }N_{i}^{4},
\end{eqnarray*}
generate a metric (solution of the Einstein equations) of type (\ref{sol12})
with arbitrary generating functions $w_{i}$ and $h_{3}$. For trivial
polarizations, when $\eta _{\alpha }\rightarrow 1$ and $\eta
_{i}^{a}\rightarrow 0,$ and $\widetilde{q}=\ ^{\circ }h_{4}=1,$ the metric (%
\ref{gsolq}) transform into the FRW metric (\ref{frw1}) if we take $a^{2}(t)$
to be determined as a solution of the Friedmann equations (\ref{fr1}) and (%
\ref{fr2}).

The goal of this section is to prove that by corresponding nonholonomic
distributions (constraints) on inhomogeneous metrics we can model de Sitter
like (exponential on $t)$ cosmological solutions of vacuum Einstein
equations. In such a model, a function $a(t)=a_{0}e^{Ht}$ contains the
Hubble constant $H$ as an experimental parameter determining a class of
nonholonomic constraints and related conformal transform with factor $q=%
\widetilde{q}(x^{k},t)a_{0}e^{Ht},$ when the equations (\ref{confeq}) are%
\begin{equation*}
\partial \ln |\widetilde{q}|/\partial x^{i}-2w_{i}H=0.
\end{equation*}%
Parametrizing $w_{i}=\ ^{t}w_{i}(t)+\ ^{s}w_{i}(x^{k})$ when $\partial _{k}\
^{s}w_{i}=\partial _{i}\ ^{s}w_{k},$ we solve this equation and satisfy the
Levi--Civita conditions (\ref{lccond12}) if
\begin{eqnarray*}
\eta _{i}^{4} &=&\ ^{1}n_{i}(x^{k}),\mbox{ \ with \ }\partial _{k}\
^{1}n_{i}=\partial _{i}\ ^{1}n_{k}, \\
\ ^{s}w_{k} &=&-\partial _{k}\ ^{\circ }h_{4} \mbox{ and } \ln |\widetilde{q}%
(x^{k},t)|=-2H\ \int \ ^{s}w_{i}(t)dx^{i}-\ ^{\circ }h_{4}(x^{i}).
\end{eqnarray*}

Putting together the above formulas in (\ref{gsolq}), we get a class of
inhomogeneous off--diagonal cosmological solutions with local anisotropy,
\begin{eqnarray}
\ ^{q}\mathbf{g} &\mathbf{=}&a_{0}^{2}e^{2Ht}\exp [-2H\ \int \
^{t}w_{i}(t)dx^{i}-\ \ ^{\circ }h_{4}(x^{i})]\mathbf{[}{dx\otimes dx}+{%
dz\otimes dz}  \notag \\
&&+\eta _{3}(x^{k},t)a_{0}^{-2}e^{-2Ht}\mathbf{e}^{3}{\otimes }\mathbf{e}%
^{3}+\ ^{\circ }h_{4}(x^{k})\mathbf{e}^{4}{\otimes }\mathbf{e}^{4}],
\label{frwan} \\
\mathbf{e}^{3} &=&dt+[\ ^{t}w_{i}(t)+\ ^{s}w_{i}(x^{k})]dx^{i},\mathbf{e}%
^{4}=dy^{4}+\ ^{1}n_{i}(x^{k})dx^{i}.  \notag
\end{eqnarray}%
This metric models a de Sitter like expansion with arbitrary generating/
integration functions ($\ ^{t}w_{i}(t),\ ^{s}w_{i}(x^{k}),\eta
_{3}(x^{k},t),\ \ ^{\circ }h_{4}(x^{i})$ $\ ^{1}n_{i}(x^{k})$) and constants
$a_{0}$ and $H$ which must be chosen following certain bounary/symmetry and
other physical superpositions to satisfy the experimental data. We can fix
such polarization functions (i.e. generating/integration functions) when a
very short accelerated ''inflationary'' stage is dominate by a locally
anisotropic and inhomogeneous vacuum solution of the Einstein equations,
lasting $\sim 10^{-36}$ and containing the de Sitter metric. Then we can say
that such an locally anisotropic stage was followed by a decelerated
homogenizing expansion, first with a radiation dominated era and then by
matter dominated era.

The class of off--diagonal inhomogeneous solutions (\ref{frwan}) is
different for the ''diagonal'' family of Szeres--Szafron metrics considered
in inhomogeneous cosmology. Here we note that the books \cite%
{kramer,krasinski}, including references within, provide a comprehensive
review of the characteristics, properties and exact and/or inhomogeneous
cosmological solutions. From any such diagonal and off--diagonal
inhomogeneous metric, we can recover the FRW model, consider solitonic
perturbations and analyze contributions of a nontrivial cosmological
constant \cite{kran}.

Our main conclusion is that using generic off--diagonal exact solutions of
the Einstein equations, with correspondingly prescribed nonholonomic
distributions, we can elaborate cosmological models with exponential
expansion and limits to FRW configurations without additional scalar fields
which would be responsible for inflation. In our approach, inflation is
modelled by nonlinear off--diagonal interactions and constraints on such a
''pure'' gravitational dynamics (in \cite{vgon}, we studied a model of
anisotropic brane inflation with off--diagonal metrics).

\section{Outlook, Discussion and Conclusions}

\label{s5}

In this work we have provided the essential features on applications in
cosmology of the anholonomic deformation method which relate the geometry of
nonholonomic manifolds to exact solutions in gravity. In particular, given
any physically important cosmological solution of Einstein equations
(defining respectively the FRW, Bianchi, Kasner, G\"{o}del or other
universes), we constructed new classes of generic off--diagonal exact
cosmological solutions which are, in general, locally anisotropic and/or
inhomogeneous. Alternately, the nonholonomic deformations described here for
cosmological solutions can be viewed as examples of the geometry of
nonholonomic distributions and generalized transforms of geometric
structures for classical and quantum (non) commutative spacetimes. While we
have considered the approach for Einstein's gravity, it is clear that the
general constructions can be extended to higher dimensions, various
Lagrange--Finsler and string/brane gravity models (albeit with increasing
computational complexity in computing off--diagonal higher dimension terms
and/or higher order nonholonomic constraints).

Our approach allows us to construct general cosmological solutions in
various gravity theories with metrics $g_{\alpha \beta }(u^{\tau })$
parametrized in the form {\footnotesize
\begin{equation*}
q^{2}\times \left|
\begin{array}{cccc}
g_{1}+\omega ^{2}(w_{1}^{\ 2}h_{3}+\omega ^{2}(n_{1}^{\ 2}h_{4}) & \omega
^{2}(w_{1}^{\ }w_{2}^{\ }h_{3}+n_{1}^{\ }n_{2}^{\ }h_{4}) & \omega ^{2}\
w_{1}^{\ }h_{3} & \omega ^{2}\ n_{1}^{\ }h_{4} \\
\omega ^{2}(w_{1}^{\ }w_{2}^{\ }h_{3}+n_{1}^{\ }n_{2}^{\ }h_{4}) &
g_{2}+\omega ^{2}(w_{2}^{\ 2}h_{3}+n_{2}^{\ 2}h_{4}) & \omega ^{2}\ w_{2}^{\
}h_{3} & \omega ^{2}\ n_{2}^{\ }h_{4} \\
\omega ^{2}\ w_{1}^{\ }h_{3} & \omega ^{2}\ w_{2}^{\ }h_{3} & h_{3} & 0 \\
\omega ^{2}\ n_{1}^{\ }h_{4} & \omega ^{2}\ n_{2}^{\ }h_{4} & 0 & h_{4}%
\end{array}%
\right|,
\end{equation*}%
} where the local coordinates are of type $u^{\tau }=(x^{i},y^{a}),$ for $%
x^{i}=(x^{3},x^{2})$ and $y^{a}=(y^{3}=t,y^{4}=y)$ and spacetime signature $%
(+,+,-,+).$ The coefficients $%
g_{k}(x^{i}),h_{a}(x^{i},t),w_{k}(x^{i},t),n_{k}(x^{i},t),q(x^{i},t)$ and $%
\omega (x^{i},t,y)$ can be defined in explicit form (following well defined
and quite simple procedures) by integrating and/or differentiating some
generating functions. Such metrics depend on certain classes of integration
functions and constants in order to define very general classes of exact
solutions in Einstein gravity and generalizations. As a matter of principle,
any solution of gravitational field equations with certain general matter
fields sources can be represented in the above generic off--diagonal form by
corresponding frame and coordinate transforms \cite{vgsolhd}. We have to
involve certain additional physical considerations, suppositions on symmetry
of interactions and boundary conditions in order to model certain realistic
cosmological models and scenaria.

An important issue which we have briefly discussed in this work concerns the
most general classes of cosmological solutions with ''nonohlonomic'' time
like variable $y^{3}=t.$ The priority of the anholonomic deformation method
is that we can determine almost all possible types of cosmological metrics
and deformations of connection in general form not making approximations
with any terms in the associated systems of partial differential equations.
The surprising propriety of the introduces nonholonomic deformations of
fundamental geometric and physical objects is that the constructions are
such way performed that we get separations of the gravitational field
equations (with respect to certain adapted frames of reference) which allows
us to generate exact solutions. Further approximations (for instance, for
generic off--diagonal metrics depending only on time variables $t$ and with
certain prescribed spacetime symmetries) are possible, but in such case
there are not lost important types of nonlinear interactions/evolutions
which can be ''lost'' during approximations for deriving effective
(simplified) systems of equations.

Furthermore, we have considered \ various generalizations of the known
important cosmological solutions which may lead to interesting new insights
into modern cosmology with locally anisotropic and inhomogeneous physical
scenaria and nonlinear interactions. These provide an interesting arena for
further exploration in gravity theories and cosmology. In principle it is
possible to model various nonstandard inflation, dark energy and dark matter
effects not introducing additional/exotic scalar and other fields but
imposing certain nonholonomic constraints on the off--diagonal dynamics of
gravitational interactions in standard Einstein gravity. In addition, it is
clear that there are many interesting directions that can be studied within
the framework of the geometry of nonholonomic distributions/frames and
off--diagonal metrics in \ gravity and cosmology. Such investigations became
possible after a general geometric \ method of constructing exact solutions
in gravity was elaborated.

\vskip5pt

\textbf{Acknowledgement: } The author is grateful to M. Anastasiei and P. Stavrinos for
important discussions.

\end{document}